\PassOptionsToPackage{unicode}{hyperref}
\PassOptionsToPackage{hyphens}{url}
\PassOptionsToPackage{dvipsnames,svgnames,x11names}{xcolor}
\documentclass[
  12pt]{article}

\usepackage{amsmath,amssymb}
\usepackage{iftex}
\ifPDFTeX
  \usepackage[T1]{fontenc}
  \usepackage[utf8]{inputenc}
  \usepackage{textcomp} 
\else 
  \usepackage{unicode-math}
  \defaultfontfeatures{Scale=MatchLowercase}
  \defaultfontfeatures[\rmfamily]{Ligatures=TeX,Scale=1}
\fi
\usepackage{lmodern}
\ifPDFTeX\else  
\fi
\IfFileExists{upquote.sty}{\usepackage{upquote}}{}
\IfFileExists{microtype.sty}{
  \usepackage[]{microtype}
  \UseMicrotypeSet[protrusion]{basicmath} 
}{}
\makeatletter
\@ifundefined{KOMAClassName}{
  \IfFileExists{parskip.sty}{%
    \usepackage{parskip}
  }{
    \setlength{\parindent}{0pt}
    \setlength{\parskip}{6pt plus 2pt minus 1pt}}
}{
  \KOMAoptions{parskip=half}}
\makeatother
\usepackage{xcolor}
\makeatletter
\ifx\paragraph\undefined\else
  \let\oldparagraph\paragraph
  \renewcommand{\paragraph}{
    \@ifstar
      \xxxParagraphStar
      \xxxParagraphNoStar
  }
  \newcommand{\xxxParagraphStar}[1]{\oldparagraph*{#1}\mbox{}}
  \newcommand{\xxxParagraphNoStar}[1]{\oldparagraph{#1}\mbox{}}
\fi
\ifx\subparagraph\undefined\else
  \let\oldsubparagraph\subparagraph
  \renewcommand{\subparagraph}{
    \@ifstar
      \xxxSubParagraphStar
      \xxxSubParagraphNoStar
  }
  \newcommand{\xxxSubParagraphStar}[1]{\oldsubparagraph*{#1}\mbox{}}
  \newcommand{\xxxSubParagraphNoStar}[1]{\oldsubparagraph{#1}\mbox{}}
\fi
\makeatother

\usepackage{longtable,booktabs,array}
\usepackage{calc} 
\usepackage{etoolbox}
\makeatletter
\patchcmd\longtable{\par}{\if@noskipsec\mbox{}\fi\par}{}{}
\makeatother
\IfFileExists{footnotehyper.sty}{\usepackage{footnotehyper}}{\usepackage{footnote}}
\makesavenoteenv{longtable}
\usepackage{graphicx}
\makeatletter
\def\maxwidth{\ifdim\Gin@nat@width>\linewidth\linewidth\else\Gin@nat@width\fi}
\def\maxheight{\ifdim\Gin@nat@height>\textheight\textheight\else\Gin@nat@height\fi}
\makeatother
\setkeys{Gin}{width=\maxwidth,height=\maxheight,keepaspectratio}
\makeatletter
\def\fps@figure{htbp}
\makeatother

\makeatletter
\@ifpackageloaded{caption}{}{\usepackage{caption}}
\AtBeginDocument{%
\ifdefined\contentsname
  \renewcommand*\contentsname{Table of contents}
\else
  \newcommand\contentsname{Table of contents}
\fi
\ifdefined\listfigurename
  \renewcommand*\listfigurename{List of Figures}
\else
  \newcommand\listfigurename{List of Figures}
\fi
\ifdefined\listtablename
  \renewcommand*\listtablename{List of Tables}
\else
  \newcommand\listtablename{List of Tables}
\fi
\ifdefined\figurename
  \renewcommand*\figurename{Figure}
\else
  \newcommand\figurename{Figure}
\fi
\ifdefined\tablename
  \renewcommand*\tablename{Table}
\else
  \newcommand\tablename{Table}
\fi
}
\@ifpackageloaded{float}{}{\usepackage{float}}
\floatstyle{ruled}
\@ifundefined{c@chapter}{\newfloat{codelisting}{h}{lop}}{\newfloat{codelisting}{h}{lop}[chapter]}
\floatname{codelisting}{Listing}

\makeatother
\makeatletter
\@ifpackageloaded{caption}{}{\usepackage{caption}}
\@ifpackageloaded{subcaption}{}{\usepackage{subcaption}}
\makeatother

\ifLuaTeX
  \usepackage{selnolig}  
\fi
\usepackage[square]{natbib}
\usepackage{bookmark}

\IfFileExists{xurl.sty}{\usepackage{xurl}}{} 
\hypersetup{
  pdftitle={Title},
  pdfauthor={Author 1; Author 2},
  pdfkeywords={3 to 6 keywords, that do not appear in the title},
  colorlinks=true,
  linkcolor={blue},
  filecolor={Maroon},
  citecolor={Blue},
  urlcolor={Blue},
  pdfcreator={LaTeX via pandoc}}

\newcommand{\anon}{1}

\usepackage{macro}
\newcommand{\Factor}{\boldsymbol{\Theta}}

\newcommand{\Loading}{\boldsymbol{\Phi}}

\newcommand{\U}{\mU}
\newcommand{\V}{\mV}
\usepackage{wrapfig}
\newtheorem*{theorem*}{Theorem}

\date{}
\begin{document}

\def\spacingset#1{\renewcommand{\baselinestretch}%
{#1}\small\normalsize} \spacingset{1}


\if1\anon
{
  \title{\bf Representation Learning with Blockwise Missingness and Signal Heterogeneity}
  \author{
    Ziqi Liu
    \\
    Department of Statistics \& Data Science, Carnegie Mellon University\\
    and \\
    Ye Tian \\
    Department of Biostatistics, Yale University \\
    and \\
    Weijing Tang \\
    Department of Statistics \& Data Science, Carnegie Mellon University
    }
  \maketitle
} \fi

\if0\anon
{
  \bigskip
  \bigskip
  \bigskip
  \begin{center}
    {\LARGE\bf Representation Learning with Blockwise Missingness and Signal Heterogeneity}
\end{center}
  \medskip
} \fi

\bigskip

\begin{abstract}
    Unified representation learning for multi-source data integration faces two important challenges: blockwise missingness and blockwise signal heterogeneity. 
    The former arises from sources observing different, yet potentially overlapping, feature sets, while the latter involves varying signal strengths across subject groups and feature sets. 
    While existing methods perform well with fully observed data or uniform signal strength, their performance degenerates when these two challenges coincide, which is common in practice. 
    To address this, we propose Anchor Projected Principal Component Analysis (APPCA), a general framework for representation learning with structured blockwise missingness that is robust to signal heterogeneity. APPCA first recovers robust group-specific column spaces using all observed feature sets, and then aligns them by projecting shared ``anchor'' features onto these subspaces before performing PCA. This projection step induces a significant denoising effect. We establish estimation error bounds for embedding reconstruction through a fine-grained perturbation analysis. In particular, using a novel \textit{spectral slicing technique}, our bound eliminates the standard dependency on the signal strength of subject embeddings, relying instead solely on the signal strength of integrated feature sets. We validate the proposed method through extensive simulation studies and an application to multimodal single-cell sequencing data.
\end{abstract}

\noindent%
\vfill

\newpage
\spacingset{1.8} 

\setlength{\abovedisplayskip}{2.2pt}
\setlength{\belowdisplayskip}{2.2pt}

\section{Introduction} \label{sec:intro}

Multi-source data integration is an important task in analyzing large-scale data collection comprised of distinct datasets~\citep{Cai2016-vw, Zhou2023-wb, Zheng2025-fl, Xue2021-sq, Xue2021-ae, Zhang2020-zc, Yu2020-ld, Chang2024-bq, Sui2025-ry}. 
However, integrating heterogeneous data sources often faces two critical challenges. The first is \emph{structured blockwise missingness}, which arises when datasets originate from different sources that measure only \emph{partially} overlapping sets of features or modalities (see Figure~\ref{fig:illustration_blockwise_missing} for an example). The second, often overlooked but equally critical, is \emph{blockwise signal heterogeneity}, where signal strengths may vary substantially across subject groups and feature sets.

These two challenges are ubiquitous and often co-occur in domains such as healthcare and single-cell genomics. For example, in multi-institutional electronic health record data, different healthcare systems often adopt different medical coding systems and collect different sets of clinical variables~\citep{Hripcsak2013-hq}. Despite substantial efforts to unify these systems, many site-specific codes still cannot be mapped to universal medical language system. Even within a single healthcare system, coding systems may evolve over time~\citep{Shi2021-bu}, so longitudinal studies naturally involve samples with only partially overlapping feature sets. Furthermore, disease-specific signals may be strong in patient subgroups but weak in healthy controls, which leads to blockwise signal heterogeneity. Another example arises in single-cell studies, where experimental design often produces datasets with different data modalities, such as gene expression, chromatin accessibility, and cell surface protein levels~\citep{Stuart2019-st, Hao2021-bs}. Here, different modalities capture different aspects of the latent state with varying strength; for instance, scRNA-seq may provide lower resolution for certain regulatory programs compared to chromatin-accessibility measurements.

In this paper, we study \emph{unified representation learning}: learning globally aligned low-dimensional subject embeddings from heterogeneous sources. Such representations provide a common latent space that supports downstream tasks such as clustering and transfer learning  when different subject groups observe only partially overlapping sets of features. However, when blockwise missingness co-occurs with blockwise signal heterogeneity, which is common, existing methods face fundamental challenges. Under blockwise missingness, information shared across subject groups is restricted to a limited set of common features.
When the signal strength in this shared block is weak, approaches that rely only on the shared features discard informative group-specific blocks and become bottlenecked by the weak shared signal~\citep{Bai2023-hj, Choi2025-lp}. Alternatively, one may adopt a two-step alignment strategy: (i)  estimate groupwise subject and feature embeddings using all  feature sets observed within each group; (ii) align the subject embeddings across groups by learning a transformation based on the feature embeddings of the shared (anchor) feature set. Existing two-step alignment methods~\citep{Zheng2024-ud, Zheng2025-fl} can work well when the within-group feature embeddings for the anchor set are accurately estimated. Under blockwise signal heterogeneity, however, these feature embeddings can become unstable and poorly estimated when the subject signals are weak within certain groups. Errors introduced in this stage contaminate the anchor-based alignment and propagate to the final globally aligned subject embeddings. It is therefore desirable to develop an alignment procedure that effectively leverages strong group-specific feature blocks while avoiding the fragility of directly matching noisy feature embeddings.

We propose \textit{Anchor Projected Principal Component Analysis} (APPCA) that aggregates signal across all available blocks while remaining robust to weak anchor signals during alignment. 
Although APPCA also follows a two-stage procedure, its key innovation lies in prioritizing groupwise subspace recovery over direct embedding matching in the alignment stage.
Specifically, APPCA first uses all observed feature blocks within each group to  estimate a reliable subspace for the groupwise subject representations. It then aligns groups using the shared anchor feature set. Rather than matching feature embeddings for the anchor set across groups, APPCA projects the anchor block onto each group’s estimated subspace and performs global representation recovery on the projected submatrix. Even when the anchor block is weak, this projection induces a substantial denoising effect by restricting high-dimensional noise to a low-dimensional subspace. As a result, APPCA avoids directly matching noisy feature embeddings and achieves accurate estimation even under heterogeneous signals.
We further extend this idea to general blockwise missingness patterns in which no single feature block is shared across all subject groups. By constructing a sequence of overlapping super-groups and applying APPCA within each super-group, we obtain a double-anchor chain-linking procedure that yields globally aligned subject representations while retaining robustness to heterogeneous signal strengths.

We also establish estimation error bounds for APPCA and its chain-linked extension, which provide theoretical insight into the gain contributed by each stage of the algorithm. 
Reflecting the two-stage structure of the algorithms, our error bound naturally decomposes into: (i) groupwise subspace recovery that leverages all feature blocks observed within each group, and (ii) global alignment that operates on the shared anchor block after projection. 
In the first stage, the error bound for groupwise subspace recovery does not rely primarily on the standalone subject signal strength. As a result, the method remains robust even when the groupwise subject signal is weak.  
In the second stage, our result further reflects that projecting the anchor block onto the aggregated groupwise subspaces fundamentally reshapes its effective signal-to-noise structure, which enables accurate global recovery even when the anchor feature signal is weak.

In particular, the robustness of groupwise subspace recovery is enabled by a new representation reconstruction error bound that remains sharp under heterogeneous signal strengths. 
In contrast to using standard perturbation arguments for subspace estimation, which leads to undesirable dependence on the condition number of the subject embeddings, the proposed bound avoids such dependence through a \textit{spectral slicing technique} and remains well controlled even when subject embeddings are ill-conditioned.
This general reconstruction error bound may be of independent interest and plays a central role in our theoretical analysis.

The rest of this paper is organized as follows. Section~\ref{sec:related} reviews related work. 
Section~\ref{sec:problem_setup} formalizes the problem setup and uses a $2\times 3$ block example to demonstrate the challenges of blockwise missingness and signal heterogeneity and providing intuition for our approach. 
Section~\ref{sec:method} presents the proposed method in general settings, with theoretical guarantees in Section~\ref{sec:theory}. 
We evaluate its empirical performance using simulation studies in Section~\ref{sec:simulation} and an application to single-cell data integration in Section~\ref{sec:real_data}.

\subsection{Related Work} \label{sec:related}

\noindent \textbf{Multi-source Data with Blockwise Missingness.}
Blockwise missingness is particularly challenging because entire feature blocks may be systematically unobserved in certain groups, which violates independent or uniform missingness assumptions and inducing highly structured observation patterns.
While a broad literature studies statistical learning problems under blockwise missingness, including regression~\citep{Kundu2019-ku, Yu2020-ld, Xue2021-sq, Xue2021-ae, Jin2023-df, Song2024-er, Xu2025-jr}, matrix completion~\citep{Cai2016-vw, Bishop2014-zj, Zhou2023-wb, Zheng2025-fl}, multitask learning~\citep{Sui2025-ry}, and clustering~\citep{Zheng2024-ud}, the problem of unified representation learning across heterogeneous feature blocks remains relatively underexplored. 
Related methods such as SPSMC~\citep{Bishop2014-zj} and BONMI~\citep{Zhou2023-wb} are further restricted to symmetric matrix settings and thus do not apply to the rectangular multi-block data matrices considered here.

More closely related, \citet{Zhang2020-zc} studied factor regression with blockwise missingness, but assumed that at least one subject group observes all features and did not account for potential blockwise signal heterogeneity. The most relevant prior works are cluster quilting~\citep{Zheng2024-ud} and Chain-linked Multiple Matrix Integration (CMMI)~\citep{Zheng2025-fl}, which can be viewed as two-step alignment methods for representation learning. Both assume relatively uniform signal strength across blocks; as we show later, their alignment step can be unstable when signal strengths are heterogeneous, particularly when some groups or blocks are weak.

\noindent \textbf{Factor Models with Weak Loadings.} 
Our work also connects to the literature on factor models in low signal-to-noise regimes. Existing work establishes the asymptotic behavior of factor estimation when loadings are weak or local-to-zero~\citep{Onatski2012-xz, Freyaldenhoven2022-yj, Uematsu2023-ng, Bai2023-hj, Choi2025-lp}. However, they are largely established for fully observed data matrices. To the best of our knowledge, the interplay between weak factor signals and blockwise missingness remains unexplored. Our work bridges this gap by establishing recovery bounds that remain robust even when signals are weak and structurally fragmented.

\subsection{Notation}

For a matrix $\mM\in\R^{d_1\times d_2}$, let $\|\mM\|_F$ and $\|\mM\|$ denote the Frobenius and operator norms; for a vector $\vv \in \R^{d}$, let $\|\vv\|_2$ denote the Euclidean norm. 
Let $\sigma_{\min}(\mM)$ and $\sigma_{\max}(\mM)$ denote the smallest and largest singular values, and $\sigma_j(\mM)$ denote the $j$th largest. 
For index sets $\gU\subseteq[d_1]$ and $\gV\subseteq[d_2]$, let $\mM_{\gU,\gV}$ denote the submatrix with rows $\gU$ and columns $\gV$; $\mM_{\gU,\cdot}$ and $\mM_{\cdot,\gV}$ restrict only rows or only columns, and we write $\mM_{\gU}:=\mM_{\gU,\cdot}$ when clear from context.
For $\mU\in\R^{d\times r}$ with rank $r$, let $\mP_{\mU}:=\mU(\mU^\top \mU)^{-1}\mU^\top$ denote the projection matrix onto the column space of $\mU$. 
We write $a_n\lesssim b_n$ or $a_n = \mathcal{O}(b_n)$ if $a_n\le C b_n$ for a universal constant $C>0$, and we write $a_n\asymp b_n$ if both $a_n\lesssim b_n$ and $b_n\lesssim a_n$.

\section{Problem Setup} \label{sec:problem_setup}

\begin{figure}[!htbp]
    \centering
    \includegraphics[width=0.5\linewidth]{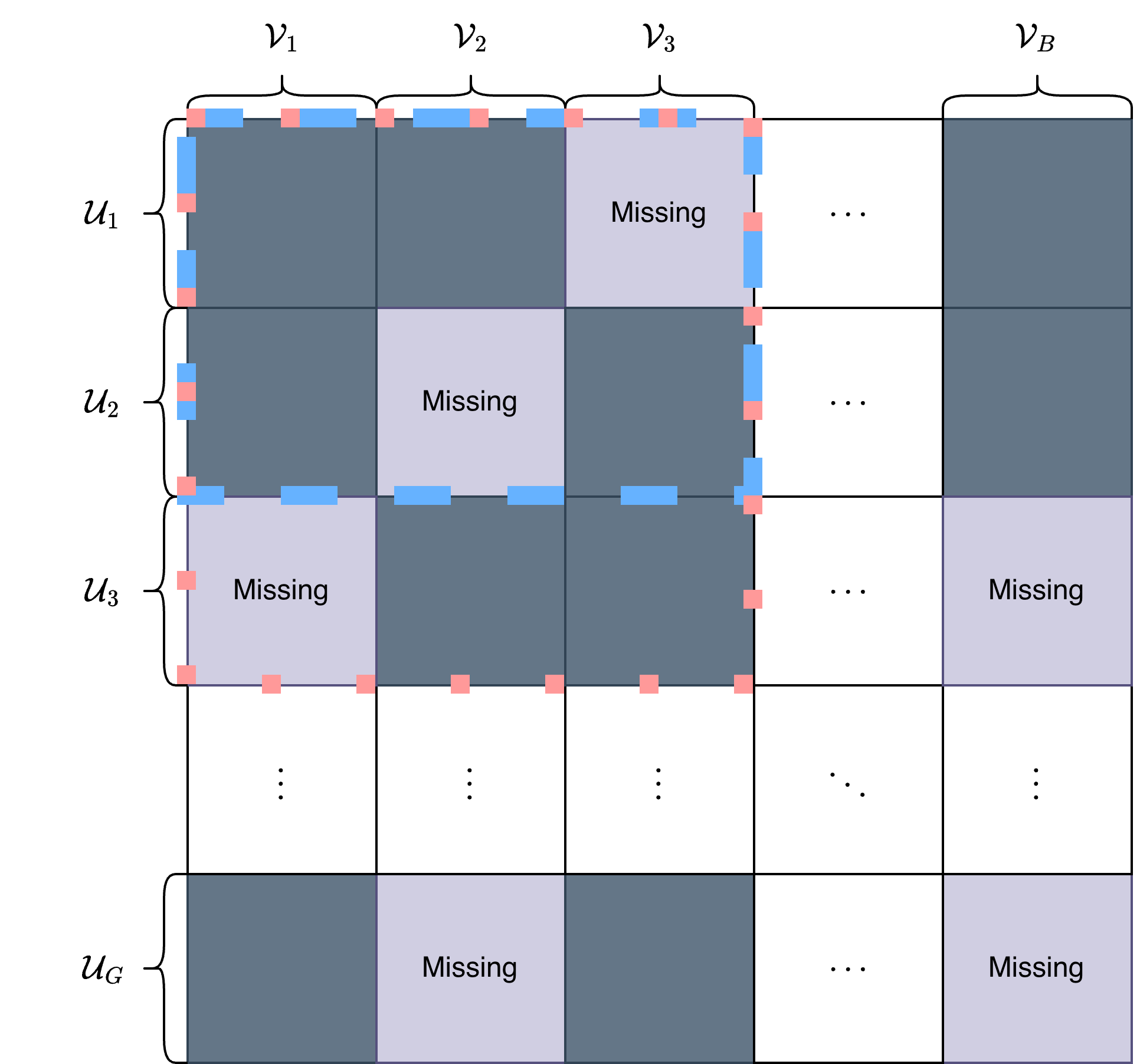}
    \caption{Schematic of the blockwise missingness pattern. Rows correspond to subject groups and columns to feature blocks. Dark regions represent observed data, whereas light regions indicate missing blocks. The area enclosed by the blue dashed line represents the $2 \times 3$ scenario analyzed in Section~\ref{subsec:challenge} and the simulation study. The region within the red dotted line illustrates the $3 \times 3$ configuration also evaluated in the simulations.}
    \label{fig:illustration_blockwise_missing}
\end{figure}

Let $\rmX$ denote the full data matrix for $n$ subjects and $p$ features. We consider the canonical model, where the data matrix $\rmX$ is modeled as a rank-$r$ matrix corrupted by additive noise:
\begin{equation}\label{eq:model}
    \rmX = \Factor \Loading^\top + \rmE,
\end{equation}
where $\Factor \in \R^{n \times r}$ and $\Loading \in \R^{p \times r}$ denote the low-dimensional representations for subjects and features, respectively, and  $\rmE\in \R^{n \times p}$ is a noise matrix with independent mean-zero sub-Gaussian entries. We assume that the $\psi_2$-norm of each entry of $\rmE$ is bounded above by~$\tau$\footnote{For a sub-Gaussian variable $X$, its $\psi_2$-norm is defined by $\|X\|_{\psi_2} = \inf_{t > 0}\{\mathbb{E}\exp(X^2/t^2) \leq 2\}$.}.
Although this canonical model has been extensively studied, real-world multi-source datasets introduce two major challenges:

\noindent \textbf{(1) Structured Blockwise Missingness.} Let the subjects, indexed by $\gU:=[n]$, be divided into $G$ disjoint groups $\gU_1, \dots, \gU_G$ such that $\gU = \cup_{g=1}^G \gU_g$. Similarly, let the features, indexed by $\gV:=[p]$, be divided into $B$ disjoint feature blocks $\gV_1, \dots, \gV_B$ with $\gV = \cup_{b=1}^B \gV_b$.
In many multi-source datasets, observations are available only at specific intersections of subject groups and feature blocks. That is, each subject group $\gU_g$ typically observes only a subset of feature blocks. 
Figure~\ref{fig:illustration_blockwise_missing} illustrates this blockwise missingness pattern. For example, subjects in group $\gU_1$ observe blocks $\gV_1, \gV_2$, and $\gV_B$, but not $\gV_3$. 

\noindent \textbf{(2) Blockwise Signal Heterogeneity.} More importantly, information is rarely distributed uniformly across subject groups or feature blocks. First, subject groups may exhibit different variability structures. For example, a disease-specific biological factor may be strong in a patient subgroup but weak in healthy controls. Second, different blocks capture different aspects of the latent state. For example, in single-cell multimodal profiling, scRNA-seq may provide a weaker signal about certain regulatory programs compared to chromatin-accessibility measurements (scATAC-seq). These examples illustrate that different sub-blocks of $X$ may have substantially different signal strengths. Such heterogeneity brings challenges to unified representation learning. 

Our goal is to recover the subject representations $\Factor$ in a unified space under this challenging regime.\footnote{While our motivating applications focus on subject embeddings, the same methodology and discussion apply to feature embeddings $\Loading$ by transposing the data matrix.} Since $\Factor$ is identifiable up to an invertible linear transformation, we aim to construct an estimator $\widehat{\Factor}$ that achieves a small \textit{alignment-adjusted} estimation error, i.e., 
\begin{equation} \label{eq:representation_recovery_error}
    \|\Delta {\widehat{\Factor}}\|_F:=\min_{H \in \R^{r \times r}}\| \widehat{\Factor} H - \Factor \|_F,
\end{equation} 
where the minimization accounts for the identifiability issue of the low-rank model.  
Without loss of generality, we assume the global subject representations $\Factor$ satisfy the scaling $\sigma_{\max}(\Factor)\asymp \sigma_{\min}(\Factor)\asymp \sqrt{n}$, as any variation in signal strength can be absorbed into~$\Loading$.\footnote{Under this standard scaling, controlling $\|\Delta {\widehat{\Factor}}\|_F$ is equivalent to controlling the scaled subspace distance $\sqrt{n}\|\mP_{\Factor} - \mP_{\widehat{\Factor}} \|_F$. We use the metric $\|\Delta {\widehat{\Factor}}\|_F$ to directly quantify the reconstruction accuracy of the latent representations themselves.}

\subsection{A Motivating Case Study} \label{subsec:challenge}
To illustrate the barriers imposed by blockwise missingness and heterogeneous signal strengths, we analyze a simplified setting with a $2 \times 3$ block structure. 
As shown in the blue dashed frame in Figure~\ref{fig:illustration_blockwise_missing}, we consider two subject groups, $\gU_1$ and $\gU_2$, with sample sizes $n_1 = |\gU_1|$ and $n_2 = |\gU_2|$. We consider three feature blocks: a shared block $\gV_1$ of dimension $p_1$, observed by both groups; and two group-specific blocks $\gV_2$ of dimension $p_2$ and $\gV_3$  of dimension $p_3$, where $\gV_2$ is observed only for $\gU_1$ and $\gV_3$ only for $\gU_2$.

\noindent \textbf{Signal Strength Structure.}
We consider a regime where the shared information is \textit{weak}, but the local information is \textit{strong}. 
Specifically, the feature signal strengths~satisfy: 
\begin{equation} \label{eq:toy_example_def_beta}
    \sigma_{\min}(\Loading_{\gV_1}^\top\Loading_{\gV_1}) \asymp p_1^\beta,
    \quad
    \sigma_{\min}(\Loading_{\gV_2}^\top\Loading_{\gV_2}) \asymp p_2, 
    \quad
    \sigma_{\min}(\Loading_{\gV_3}^\top\Loading_{\gV_3}) \asymp p_3, \ \text{for some } 0 < \beta < 1.
\end{equation}
Regarding the subject signal strengths, we consider weak signals within each group:
\begin{equation} \label{eq:toy_example_def_alpha}
    \sigma_{\min}(\Factor_{\gU_1}^\top \Factor_{\gU_1}) \asymp n_1^{\alpha},
    \qquad
    \sigma_{\min}(\Factor_{\gU_2}^\top \Factor_{\gU_2}) \asymp n_2^{\alpha}, \quad \text{for some } 0 < \alpha < 1.
\end{equation}
For ease of exposition, we consider balanced dimensions, with $n_1 \asymp n_2 \asymp n$ and $p_1 \asymp p_2 \asymp p_3 \asymp p$, and assume that the global signal scales as $\sigma_{\min}(\Factor^\top \Factor) \asymp n$ and $\sigma_{\min}(\Loading^\top \Loading) \asymp p$. This implies that, if all blocks were fully observed, the optimal estimation rate would be achievable. With blockwise missingness, however, this setup creates a challenging regime that exposes the limitations of existing approaches. Next, we review several baseline methods and discuss their limitations within this setup. 

\noindent \textbf{Baseline 1: Using Shared Feature Block.} 
A simple approach is to discard group-specific blocks and perform PCA solely on the shared submatrix $X_{\gU, \gV_1}$. This reduces the problem to  factor estimation with weak loadings \citep{Bai2023-hj, Choi2025-lp}. Let $\widehat{\Factor}_{\text{shared}}$ denote the estimator obtained by taking the top-$r$ left singular vectors of this submatrix and scaling them by $\sqrt{n}$, then the estimation error satisfies that, with high probability,
\begin{equation} \label{eq:bai_ng_bound}
    \frac{1}{\sqrt{n}} \|\Delta \widehat{\Factor}_{\text{shared}} \|_F
    \;\lesssim\;
    \frac{1}{p^{\beta/2}} + \frac{p}{n p^\beta}.
\end{equation}
This bound suggests that the estimation accuracy is bottlenecked by the weak signal parameter $\beta$ of the shared block. In particular, in the regime where $p \lesssim n$, the error scales as $\mathcal{O}(p^{-\beta/2})$, which is substantially slower than the rate $\mathcal{O}(p^{-1/2})$ achievable using the full matrix with strong signals. Consequently, by failing to leverage the strong signals in the group-specific blocks $\gV_2$ and $\gV_3$, this approach results in suboptimal recovery.

\noindent \textbf{Baseline 2: Two-step Embedding Alignment.}
To leverage the strong signals in the group-specific blocks $\gV_2$ and $\gV_3$, one may alternatively adopt a ``local embedding alignment'' strategy. Adapting the CMMI framework \citep{Zheng2025-fl} to our asymmetric context includes two steps. First, we compute local estimators on the available data for each group. Specifically, we obtain $\widehat{\Factor}_{\gU_1}^{(1)}, \widehat{\Loading}_{\gV_1 \cup \gV_2}^{(1)}$ from $X_{\gU_1, \gV_1 \cup \gV_2}$ via PCA and similarly obtain $\widehat{\Factor}_{\gU_2}^{(2)}, \widehat{\Loading}_{\gV_1 \cup \gV_3}^{(2)}$ from $X_{\gU_2, \gV_1 \cup \gV_3}$.
Second, we learn a linear transformation $\widehat{\mW}$ to align the coordinate systems by minimizing the difference between the embeddings for the shared feature block: $\widehat{\mW} = \argmin_{\mW \in \R^{r\times r}} \left\| \widehat{\Loading}_{\gV_1}^{(1)} \mW^\top - \widehat{\Loading}_{\gV_1}^{(2)} \right\|_F$. 
The final estimator is constructed by concatenating the aligned representations: $\widehat{\Factor}_{\text{CMMI}} = \left[\left(\widehat{\Factor}_{\gU_1}^{(1)}\right)^\top, \left(\widehat{\Factor}_{\gU_2}^{(2)}\widehat{\mW}\right)^\top \right]^\top $.
Consequently, the total error comes from two sources: the local estimation error and the cross-group alignment error. 
While the local estimation step benefits from the stronger signals in $\gV_2$ and $\gV_3$,
the method is vulnerable during the alignment step. This is because the transformation $\widehat{W}$ relies on the estimated feature embeddings for the shared block $\gV_1$. When the subject signals in $\gU_1$ are weak or sample sizes are insufficient relative to $p$, the estimation of the feature embeddings $ \widehat{\Loading}_{\gV_1}^{(1)}$ becomes inaccurate. This leads to a poorly estimated transformation matrix $\widehat{W}$, which propagates error into the final estimator.

Overall, this simple example reveals caveats of existing methods under blockwise missingness and signal heterogeneity. The first method that relies on the shared feature block avoids alignment issues but is  constrained by the weak signal strength in the shared feature block. In contrast, the two-step alignment method effectively leverages strong signals within each group but suffers from the alignment error when the subject signals are weak within groups. 
These observations motivate our proposed method that uses strong signals in group-specific blocks while avoiding the embedding alignment~issue.

\subsection{Overview of Proposed Approach: Anchor Projected PCA} \label{subsec:projected_pca_toy_example}

We now introduce our method, \textit{Anchor Projected PCA} (APPCA), under this $2\times 3$ block setting and provide an informal theoretical guarantee. 
This example serves to explain the core ideas behind our method.
A full presentation of the method for general blockwise missingness, along with its formal theoretical results, will be given in Sections~\ref{sec:method} and~\ref{sec:theory},~respectively. 

APPCA leverages strong signals in group-specific feature blocks while maintaining robustness to heterogeneous signal strengths. Compared to the two-step alignment method, which aligns the estimated feature embeddings for the shared block directly, our method projects the shared feature block onto the estimated subspaces. We refer to this as ``Anchor Projected'' PCA because the shared feature block $\gV_1$ acts as a statistical \textit{anchor} that bridges the disjoint subject groups. The method consists of two stages:

\noindent \textbf{Stage 1: Groupwise Subspace Estimation.}
    We leverage strong signals in group-specific feature blocks to estimate the column space for each subject group.
    Let $\hat{\U}_{\gU_1}^{(1)}$ and~$\hat{\U}_{\gU_2}^{(2)}$ denote the top-$r$ left singular vectors of the sub-matrices $\rmX_{\gU_1, \gV_1 \cup \gV_2}$ and $\rmX_{\gU_2, \gV_1 \cup \gV_3}$,~respectively. 
    
\noindent \textbf{Stage 2: Global Alignment via Anchor Projection.}
    We denoise the shared ``anchor'' block by projecting it onto the subspaces estimated in Stage 1:
    \begin{equation*}
        \widehat{\rmX}_{\text{proj}} = \begin{bmatrix}
            \mP_{\hat{\U}_{\gU_1}^{(1)}} \rmX_{\gU_1, \gV_1} \\
            \mP_{\hat{\U}_{\gU_2}^{(2)}} \rmX_{\gU_2, \gV_1}
        \end{bmatrix},
    \end{equation*}
    where $\mP_M$ denotes the projection matrix associated with the column space of $\mM$. 
    The global subject representation $\widehat{\Factor}_{\text{APPCA}}$ is obtained by computing the top-$r$ left singular vectors of $\widehat{\rmX}_{\text{proj}}$ and scaling by $\sqrt{n}$.

We characterize the performance of $\widehat{\Factor}_{\text{APPCA}}$ in this $2 \times 3$ block setting by specializing our general theory in Section~\ref{sec:theory} to the signal structure described above. The result is presented in the following informal theorem.
\begin{theorem*}[Informal] With high probability, the Anchor Projected PCA satisfies:
    \begin{equation} \label{eq:projected_pca_bound_toy_example}
        \frac{1}{\sqrt{n}} \|\Delta \widehat{\Factor}_{\text{APPCA}}\|_F  
        \;\lesssim\; 
        \underbrace{\frac{1}{\sqrt{p}} + \frac{p}{n p^\beta}}_{\text{Dominant Terms}} 
        + \underbrace{\frac{1}{\sqrt{n^{1+\alpha}}}\left(1 + \frac{n}{p}\right) + \frac{\log n}{\sqrt{np^\beta}}}_{\text{Smaller-order Terms}}.
    \end{equation}
\end{theorem*}
For ease of comparison, consider the high-dimensional regime where the sample size $n$ and feature dimension $p$ grow proportionally, i.e., $n \asymp p$. In this regime, the error bound simplifies to $\|\Delta \widehat{\Factor}_{\text{APPCA}}\|_F /\sqrt{n} \lesssim p^{-1/2} + p^{-\beta}$.
This result implies two critical improvements over the baselines, enabled by the \textit{subspace denoising} inherent in the projection step.

\noindent \textbf{(1) Robustness to Weak Block Signals (vs. Shared-Block PCA):} The first dominant error term improves from $\mathcal{O}(p^{-\beta/2})$ in~\eqref{eq:bai_ng_bound} to $\mathcal{O}(p^{-1/2})$. This improvement occurs because APPCA breaks the information bottleneck imposed by the weak shared block. By leveraging the strong signals in the group-specific blocks $\gV_2$ and $\gV_3$, we estimate the groupwise subspaces $\hat{\U}_{\gU_1}^{(1)}$ and $\hat{\U}_{\gU_2}^{(2)}$ with high accuracy.

\noindent \textbf{(2) Robustness to Weak Group Signals (vs. Two-Step Embedding Alignment):} Importantly, our bound eliminates the impact of weak group-level signals. The term involving group-wise signal strengths ($n^{-\alpha/2}$) appears only in smaller-order terms scaled by a factor of $n^{-1/2}$, which becomes negligible as the sample size increases. 
This robustness is achieved because the projection $\mP_{\hat{\U}^{(g)}}$ acts as a denoising filter on the shared anchor block. When applied to $\rmX_{\cdot, \gV_1}$, it reduces noise while preserving the underlying structure:
\begin{equation*}
    \mP_{\hat{\U}^{(g)}} \rmX_{\gU_g, \gV_1} \approx \mP_{\U_{\gU_g}} (\Factor_{\gU_g} \Loading_{\gV_1}^\top + \rmE) = \Factor_{\gU_g} \Loading_{\gV_1}^\top + \mP_{\U_{\gU_g}}\rmE.
\end{equation*}
\begin{wrapfigure}{r}{0.5\textwidth}
    \vspace{-5mm}
    \begin{center}
        \includegraphics[width=1\linewidth]{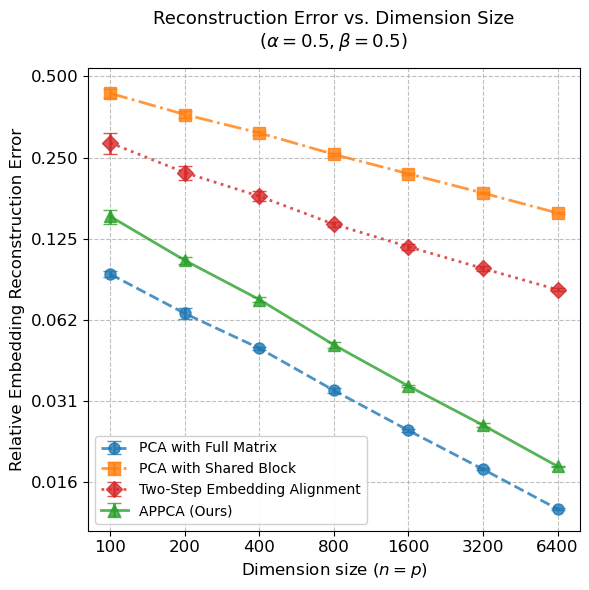}
    \end{center}
    \caption{Log-log plot of the estimation error of $\Factor$ with respect to dimension $n=p$.}
    \label{fig:sim_2_3_sim}
    \vspace{-5mm}
\end{wrapfigure}
Since the noise matrix $\rmE$ is projected onto a low-dimensional subspace ($r \ll n$), its effective variance is greatly reduced. This ``denoised'' global matrix allows for effective information pooling across all $n$ subjects, which mitigates the alignment instability in the two-step embedding alignment. 

We further demonstrate the effect of the anchored projection step with a numerical experiment under the setting $\alpha=\beta=0.5$ and $n=p$.
Figure~\ref{fig:sim_2_3_sim} reports the estimation error on a log–log scale. The two baselines exhibit a slower convergence rate (indicated by a smaller slope) than our method. Notably, our method aligns closely with the oracle rate obtained when the full matrix is observed. Full details about the simulation studies are provided in Section~\ref{sec:simulation}. 

\section{Methodology}\label{sec:method}

In this section, we develop a general framework for recovering global subject embeddings under blockwise missingness and signal heterogeneity. 
Our approach builds on the insights gained from the simple $2\times 3$ block example and extends it to general blockwise missingness structures.
Let $\mM \in \{0,1\}^{G \times B}$ denote the indicator matrix for blockwise missingness, where $\mM_{g,b} = 1$ if subject group $g$ observes feature block $b$, and $0$ otherwise. 
For each subject group $g \in \{1, \dots, G\}$, we define the set of observed features as the union of all observed blocks:
\begin{equation} \label{eq:def_mathcal_V_tuple_g}
    \gV_{(g)} = \bigcup_{b: \mM_{g,b}=1} \gV_b.
\end{equation}
We adopt a modular estimation strategy. 
First, in Section~\ref{subsec:shared_feature}, we consider the base case where all groups share at least one common feature block. We introduce \textit{Anchor Projected PCA}, which serves as the \textit{core building block} for our general procedure. Next, in Section~\ref{subsec:chain_link}, we address the general blockwise missingness setting by decomposing the global recovery problem into a sequence of overlapping local subproblems. By iteratively applying the core method in Section~\ref{subsec:shared_feature} and linking the estimates through a chain-linking strategy, we obtain globally aligned subject embeddings.

\subsection{The Base Case: When a Shared Feature Set Exists}\label{subsec:shared_feature}
We first consider the scenario where all subject groups observe at least one common feature block. We define the \textit{shared feature set} $\gT$ as the intersection of all group-specific feature~sets:
\begin{equation} \label{eq:def_mathcal_T}
    \gT: = \bigcap_{g=1}^G \gV_{(g)} = \bigcup_{b: \forall g, M_{g,b} = 1} \gV_b, \qquad \text{with }\gT \neq \emptyset.
\end{equation}
In this regime, we generalize the strategy introduced in Section~\ref{subsec:projected_pca_toy_example} and summarize APPCA in Algorithm~\ref{alg:projected_pca_full_observe}. The method consists of two stages: (1) groupwise subspace estimation, where we estimate the column space for each subject group using the maximal set of observed features, $\gV_{(g)}$; and (2) alignment via anchor projection, where we project the shared feature block $X_{\cdot, \gT}$ onto the estimated groupwise subspaces and obtain a globally aligned representation.
\begin{algorithm}[!htbp]
    \caption{Anchor Projected PCA (When a Shared Feature Set Exists)}
    \label{alg:projected_pca_full_observe}
    \KwIn{Data $\rmX$, groups $\{\gU_g\}_{g=1}^G$, feature blocks $\{\gV_b\}_{b=1}^B$, indicator $\mM$, rank $r$.}
    \KwOut{Estimated factor matrix $\hat{\Factor} \in \R^{n \times r}$.}
    
    \Comment{Initialization}
    Identify the group-specific feature sets $\gV_{(g)}$ via~\eqref{eq:def_mathcal_V_tuple_g} and the shared set $\gT$ via~\eqref{eq:def_mathcal_T}. \;
    
    \Comment{Stage 1: Groupwise Subspace Estimation}
    \For{$g = 1$ \KwTo $G$}{
        Compute $\hat{\U}_{\gU_g}^{(g)} \gets \operatorname{SVD}_r(\rmX_{\gU_g,\gV_{(g)}})$ \Comment{top-$r$ left singular vectors}
    }    
    \Comment{Stage 2: Global Alignment via Anchor Projection}
    Construct the projected data matrix using the shared feature set:
    \[
      \tilde{\rmX} =
      \begin{bmatrix}
        \mP_{\hat{\U}_{\gU_1}^{(1)}} \rmX_{\gU_1, \gT} \\
        \vdots \\
        \mP_{\hat{\U}_{\gU_G}^{(G)}} \rmX_{\gU_G, \gT}
      \end{bmatrix},
      \quad \text{where } \mP_{\hat{\U}_{\gU_g}^{(g)}} = \hat{\U}_{\gU_g}^{(g)} \hat{\U}_{\gU_g}^{(g)\top}.
    \]
    Compute $\hat{\U} \in \R^{n \times r}$ as the top-$r$ left singular vectors of $\tilde{\rmX}$.\;
    \KwRet $\hat{\Factor} \gets \sqrt{n} \hat{\U}$ \;
\end{algorithm}

The rationale for Algorithm~\ref{alg:projected_pca_full_observe} builds directly on the insights from the $2\times 3$ block example, where subspace denoising via projection plays an important role. Learning subject representations based solely on the shared features $\rmX_{\cdot, \gT}$ may suffer from weak feature signals. To address this, Stage 1 of our method first constructs accurate groupwise subspace estimates by leveraging the strong, group-specific feature blocks available to each group.
In contrast to the two-step embedding alignment method \citep{Zheng2025-fl}, which relies on accurate recovery of feature embeddings and thus degrades when group subject signals are weak, Stage 2 of our method avoids this by projecting $\rmX_{\cdot, \gT}$ onto these accurately estimated subspaces. This projection step allows the final PCA step to effectively recover the global subject embeddings even under blockwise signal heterogeneity.

\subsection{The General Case: Sequential Chain Linking} \label{subsec:chain_link}

In the general blockwise missingness setting, a single globally shared feature block may not exist (i.e., $\gT = \varnothing$). Nevertheless, global recovery remains feasible provided that subject groups are linked through overlapping memberships. We exploit this structure through a \textit{double-anchor chain-linking} strategy,  which constructs globally aligned subject representations by sequentially estimating and aligning locally anchored subspaces. 

Specifically, we assume there exists a sequence of \textit{super-groups} $\gS_1, \dots, \gS_K$, where each $\gS_k \subseteq \gU$ is a union of several subject groups, satisfying the following conditions:
\begin{enumerate}
    \item \textbf{Super-Group Anchoring:} Within each super-group $\gS_k$, all subjects observe at least one common feature block. This shared block serves as an \textit{inner anchor} so that Algorithm~\ref{alg:projected_pca_full_observe} can be applied to obtain a local subspace estimate.
    \item \textbf{Sequential Overlap:}  Each super-group $\gS_{k}$ has non-empty intersection with the union of all previous super-groups, i.e., $\gO_k := (\bigcup_{j=1}^{k-1} \gS_j) \cap \gS_{k} \neq \varnothing$ for all $k = 2, \dots, K$. These overlapping subjects act as \textit{outer anchors} for aligning estimated subspaces across super-groups.
    \item \textbf{Global Coverage:} The union of the super-group sequence spans all subjects, i.e., $\bigcup_{k=1}^{K} \gS_k = \gU$.
\end{enumerate}

Given such a sequence, we introduce Algorithm~\ref{alg:projected_pca_chain_link}, which sequentially estimates, aligns, and concatenates the subspaces across super-groups.
Specifically, for each super-group $\gS_k$, we apply APPCA in Algorithm~\ref{alg:projected_pca_full_observe} to obtain a robust estimate of the super-group column space.  Importantly, this estimate inherits robustness to heterogeneous signal strengths across both feature blocks and subject groups through anchor projection. This step differs fundamentally from the asymmetric version of the CMMI procedure in~\citet{Zheng2025-fl} and the cluster quilting procedure in~\citet{Zheng2024-ud}, which align feature embeddings that are themselves estimated under potentially weak subject signals. In contrast, each super-group subspace in our method is first stabilized by an inner-anchor projection before any cross-group alignment is performed. 

We then align adjacent super-group subspaces by solving for the optimal linear transformation that best matches the representations of their overlapping subjects. Iterating this alignment step along the sequence generates globally aligned embeddings for all subjects.

\begin{algorithm}[!htbp]
    \caption{Double Anchor Projected PCA via Chain Linking}
    \label{alg:projected_pca_chain_link}
    \KwIn{Data matrix $X$, chain sequence $\{\gS_k\}_{k=1}^K$, target rank $r$.}
    \KwOut{Global factor estimator $\hat{\Factor} \in \R^{n \times r}$.}
    
    \Comment{Step 1: Initialization}
    Compute $\hat{\Factor}_{\gS_1}$ via Algorithm~\ref{alg:projected_pca_full_observe} on $X_{\gS_1, \gV}$; 
    set $\hat{\U}\gets \text{orthonormal basis of }\hat{\Factor}_{\gS_1}$. \;
    
    \Comment{Step 2: Sequential Alignment and Concatenation}
    \For{$k = 2$ \KwTo $K$}{
        Compute $\tilde{\Factor}_{\gS_k}$ via Algorithm~\ref{alg:projected_pca_full_observe} on $X_{\gS_k, \gV}$; 
        set $\tilde{\U}_{\gS_k} \gets \text{orthonormal basis of }(\tilde{\Factor}_{\gS_k})$. \;
        
        $\gO_k \gets (\bigcup_{j=1}^{k-1} \gS_j) \cap \gS_k,\quad \gR_k \gets \gS_k\setminus (\bigcup_{j=1}^{k-1} \gS_j)$. \;
        
        $\mA \gets \hat{\U}_{\gO_k},\quad \mB \gets (\tilde{\U}_{\gS_{k}})_{\gO_k}$. \;
        $\mW_k \gets (\mB^\top \mB)^{-1}\mB^\top \mA$. \Comment{optimal linear transformation from $\mB$ to $\mA$. }
        $\hat{\U}\gets 
        \begin{bmatrix}
            \hat{\U}\\
            (\tilde{\U}_{\gS_{k}})_{\gR_k}\mW_k
        \end{bmatrix}$;\;
        $\hat{\U} \gets \text{orthonormal basis of }\hat{\U}$. \;
    }
    \KwRet $\hat{\Factor} \gets \sqrt{n}\hat{\U}$. \;
\end{algorithm}

\section{Theoretical Results} \label{sec:theory}

\subsection{Estimation Error Bound for Algorithm~\ref{alg:projected_pca_full_observe}}
In Algorithm~\ref{alg:projected_pca_full_observe}, the projection matrix $\mP$ constructed in Stage~2 is estimated using data that overlaps with the anchor submatrix $\rmX_{\cdot,\gT}$, which induces the dependence between $\mP$ and $\rmX_{\cdot,\gT}$. For ease of theoretical analysis, we introduce a cross-fitted variant of APPCA described in Supplementary Material A, which exhibits nearly identical empirical performance. In this variant, Stage~1 and Stage~2 are performed on independent data splits, so that the projection matrix $\mP$ is independent of the anchor block $\rmX_{\cdot,\gT}$.  For notational simplicity, we continue to refer to Algorithm~\ref{alg:projected_pca_full_observe} without explicitly distinguishing the cross-fitted version. All theoretical results below should be interpreted under this independence regime.

We analyze the estimation error of the subject embedding estimator $\widehat{\Factor}$ returned by Algorithm~\ref{alg:projected_pca_full_observe}. Note that the error metric $\Delta \widehat{\Factor}$ defined in (\ref{eq:representation_recovery_error})  can be equivalently written as the residual of $\Factor$ orthogonal to the estimated column space: 
\begin{equation}\label{eq:embedding_subspace_equivalence}
    \|\Delta {\widehat{\Factor}}\|_F = \left\| (\mI - \mP_{\hat{\U}})\Factor \right\|_F,
\end{equation}
where $\hat{\U}$ is the top-$r$ left singular subspace of $\tilde{\rmX}$ in Algorithm~\ref{alg:projected_pca_full_observe}.  
The two-stage structure of Algorithm~\ref{alg:projected_pca_full_observe} leads to a natural decomposition of this error. Define the block-diagonal projection matrix $\mP \coloneqq \operatorname{diag}\left( \mP_{\hat{\U}_{\gU_1}^{(1)}}, \dots, \mP_{\hat{\U}_{\gU_G}^{(G)}} \right) \in \R^{n \times n}$ that aggregates the groupwise subspace estimates from Stage~1.
By the triangle inequality, the error decomposes~as
\begin{align}
    \|\Delta {\widehat{\Factor}}\|_F &\le \left\|(\mI - \mP_{\hat{\U}})(\Factor - \mP\Factor) \right\|_F + \left\|(\mI - \mP_{\hat{\U}})\mP\Factor \right\|_F \nonumber \\
    &\le \underbrace{\left\|\Factor - \mP\Factor \right\|_F}_{\text{(I) Aggregated Subspace Projection Error}} + \underbrace{\left\|(\mI - \mP_{\hat{\U}})\mP\Factor \right\|_F}_{\text{(II) Global PCA Error on the Projected Subspace}}. \label{eq: decomp}
\end{align}
Term (I) captures the aggregated error from projecting $\Factor$ onto the groupwise subspaces learned in Stage~1. Term (II) is the PCA error incurred in Stage~2 when estimating the column space from the projected anchor block. 
We analyze them separately to disentangle the distinct sources of error and to clarify the gain contributed by each stage of the algorithm.  

\noindent \textbf{Groupwise Subspace Recovery in Stage~1.}
We first bound term (I) by establishing groupwise subspace recovery guarantees for each group $g\in [G]$ in Stage~1. Let $n_g = |\gU_g|$ denote the number of subjects and $p_{(g)} = |\gV_{(g)}|$ denote the number of observed features in group $g$.
We then establish the error bound on the groupwise subspace recovery $\left(\mI - \mP_{\hat{\U}_{\gU_g}^{(g)}}\right) {\Factor}_{\gU_g}$, which constitutes term (I) in the decomposition~(\ref{eq: decomp}).

\begin{restatable}{theorem}{ThmSharedFeatureSetStep} \label{thm:shared_feature_set_step_1}
    Assume the model in \eqref{eq:model}. Suppose there exists a constant $C_1$ such that 
    \begin{equation} \label{eq:assump_local_snr}
        \sigma_{\min}\left( \Factor_{\gU_g}\Loading_{\gV_{(g)}}^\top \right) \geq C_1  r \tau \left(\sqrt{n_g} + \sqrt{p_{(g)}} \right), 
    \end{equation}
    then there exist constants $c$ and $C_2>0$ such that, with probability at least $1 - n_g^{-c}$, 
    \begin{equation} 
    \label{eq:local_subspace_recovery_error_bound} 
        \left\| \left(\mI - \mP_{\hat{\U}_{\gU_g}^{(g)}}\right) \Factor_{\gU_g} \right\| \leq C_2 \frac{\mathcal{E}_g}{\sigma_{\min}(\Loading_{\gV_{(g)}})}, \ \ \text{with}\ \ \mathcal{E}_g \coloneqq r^2\tau \sqrt{n_g}  + \frac{r^2 \tau^2 \left( n_g + p_{(g)} \right)}{\sigma_{\min}(\Factor_{\gU_g}\Loading_{\gV_{(g)}}^\top)}.
    \end{equation}
\end{restatable}

Condition~\eqref{eq:assump_local_snr} is a standard assumption in spectral analysis, which requires a sufficiently large signal in the groupwise matrix $\Factor_{\gU_g}\Loading_{\gV_{(g)}}^\top$ against the sub-Gaussian noise, whose spectral norm scales as $O(\sqrt{n_g} + \sqrt{p_{(g)}})$. The proof of Theorem~\ref{thm:shared_feature_set_step_1} is provided in Supplementary Material C.1. We note that a direct application of existing perturbation bounds for column subspaces would yield an error bound that depends on the signal strength of the subject embedding~${\Factor}_{\gU_g}$ itself alone, which can be suboptimal when ${\Factor}_{\gU_g}$ is weak. 
In contrast, our result does not rely on the standalone signal strength of $\Factor_{\gU_g}$. Instead, $\Factor_{\gU_g}$ appears in the bound only through the product matrix $\Factor_{\gU_g}\Loading_{\gV_{(g)}}^\top$ in a term of a smaller-order (see Remark~\ref{rmk:local_estimation_error} for details). This refinement is enabled by a novel \textit{spectral slicing technique}. We present this technique as a general result of independent interest in Section~\ref{subsec: spectral slicing}. 

\begin{remarknum} \label{rmk:local_estimation_error} 
    By the equivalence in~(\ref{eq:embedding_subspace_equivalence}), Theorem~\ref{thm:shared_feature_set_step_1} implies a high-probability error bound for the groupwise subject embeddings $\hat{\Factor}_{\gU_g}^{(g)} = \sqrt{n_g}\hat{\U}_{\gU_g}^{(g)}$:
    \begin{equation} \label{eq:local_estimation_error_bound}
        \frac{1}{\sqrt{n_g}}\left\|\Delta\hat{\Factor}_{\gU_g}^{(g)}\right\|_F \lesssim \frac{\sqrt{r}}{\sqrt{n_g}} \left( \frac{r^2\tau \sqrt{n_g} }{\sigma_{\min}(\Loading_{\gV_{(g)}})} + \frac{r^2 \tau^2 \left( n_g + p_{(g)} \right)}{\sigma_{\min}(\Loading_{\gV_{(g)}})\sigma_{\min}(\Factor_{\gU_g}\Loading_{\gV_{(g)}}^\top)} \right).
    \end{equation}
    To gain further insight, we consider  a signal-strength regime consistent with the discussion in Section~\ref{subsec:challenge}. Suppose the signal strength in feature embeddings is strong, with $\sigma_{\min}(\Loading_{\gV_{(g)}}) \asymp p_{(g)}^{1/2}$, while that in the groupwise subject embeddings may be weak, with $\sigma_{\min}(\Factor_{\gU_g}) \asymp n_g^{\alpha_g/2}$ for some $\alpha_g \in (0,1]$. With constant rank and noise level, i.e., $r, \tau = \mathcal{O}(1)$, the bound in~\eqref{eq:local_estimation_error_bound} simplifies to $p_{(g)}^{-1/2}+ n_{(g)}^{(1-\alpha_g)/2}p_{(g)}^{-1} + n_{(g)}^{-(1+\alpha_g)/2}$.
    When $n_{(g)} \asymp p_{(g)}$, the leading term is $\mathcal{O}(p_{(g)}^{-1/2})$, yielding two important implications:
    \begin{enumerate}
        \item \textit{Benefit of Feature Integration:} The dominant error term is controlled by \(\sigma_{\min}(\Loading_{\gV_{(g)}})\). 
        By integrating all observed feature blocks for group \(g\), rather than relying solely on the shared set \(\gT\), the method exploits stronger signals in \(\gV_{(g)}\) and achieves more accurate groupwise subject embedding recovery.
    
        \item \textit{Robustness to Weak Signal in Subject Embeddings:} The signal strength in subject embeddings, captured by $\alpha_g$, affects the bound only through smaller-order terms. Thus, our method remains robust even when subject embeddings are weak. In contrast, two-step embedding alignment methods rely on accurate estimation of feature embeddings for the shared feature block \(\gT\). However, feature embedding estimation itself depends on the signal strength of \(\Factor_{\gU_g}\); consequently, weak subject signals can propagate through the alignment step and degrade the final estimation of global subject embeddings.  
    \end{enumerate}
\end{remarknum}

\noindent \textbf{Global Subject Embedding Recovery in Stage~2.} After estimating the groupwise subspaces $\mP_{\hat{\U}_{\gU_g}^{(g)}}$ in Stage~1, Stage~2 of Algorithm~\ref{alg:projected_pca_full_observe} aggregates them into the global projection operator $\mP$ and then performs PCA on the projected anchor matrix $\mP \rmX_{\cdot, \gT}$ to obtain $\hat{\U}$.
Term (II) in the decomposition~\eqref{eq: decomp} corresponds to the PCA estimation error incurred on this projected anchor block. 
Note that the projected anchor matrix can be decomposed as $\mP\Factor\Loading^\top_{\gT} + \mP \rmE_{\cdot, \gT}$.
The key insight is that projection onto $\mP$ effectively denoises the high-dimensional noise $\rmE_{\cdot, \gT}$ by restricting it to a lower-dimensional subspace, thereby substantially improving the effective signal-to-noise ratio and enabling accurate recovery of the global column space (see Remark~\ref{rmk:global_alignmnt_error} for details).
Similar concept of denoising via projection has been explored in \citet{Fan2016-af, Fan2022-kf}, where the projection operators  rely on external covariates or domain-specific knowledge. In our case, the denoising projection is learned from group-specific feature blocks.
The following theorem formalizes this intuition and establishes the error bound for the output of Algorithm~\ref{alg:projected_pca_full_observe}.

\begin{theorem} \label{thm:shared_feature_set}
    Assume the model in \eqref{eq:model}. Suppose there exists a constant $C_1$ such that Condition in~\eqref{eq:assump_local_snr} holds for all $g=1\dots, G$, and 
    \begin{equation} \label{eq:assump_global_snr}
        \sigma_{\min}(\Factor \Loading_{\gT}^\top) \geq C_1  r^2 \tau^2 \sqrt{G} \cdot \max_{g}\left\{\sqrt{n_g} + \sqrt{p_{(g)}}\right\},
    \end{equation}
    then there exist constants $c$ and $C_2>0$ such that, with probability at least $1 - (G+3)(\min_{1\le g\le G} n_g)^{-c}$, the output $\widehat{\Factor}$ of Algorithm~\ref{alg:projected_pca_full_observe}  satisfies: 
    \begin{equation} \label{eq:bound_shared_feature_set}
        \frac{1}{\sqrt{n}} \left\|\Delta\widehat{\Factor}\right\|_F \le \frac{C_2\sqrt{r}}{\sqrt{n}} \left( \sqrt{\sum_{g=1}^G \left(\frac{ \mathcal{E}_g }{\sigma_{\min}(\Loading_{\gV_{(g)}})} \right)^2} + \frac{ \Xi(\gT) + \log n}{\sigma_{\min}(\Loading_{\gT})} \right),
    \end{equation}
    where $n = \sum_{g=1}^G n_g$, $\mathcal{E}_g$ is defined in~\eqref{eq:local_subspace_recovery_error_bound}, and 
    \begin{equation} \label{eq:def_Xi}
        \Xi(\gT) \coloneqq r^2 \tau \sqrt{Gr} + \frac{r^2\tau^2 \left( Gr + |\gT| \right)}{\sigma_{\min}(\Factor\Loading_{\gT}^\top)}.
    \end{equation}
\end{theorem}

Condition~\eqref{eq:assump_global_snr} requires that the signal strength in the shared anchor block be sufficient to dominate both the noise and the  subspace estimation errors accumulated in Stage~1. In a balanced regime where the block/group sizes are comparable (i.e., $n_g \asymp n$, $p_{g} \asymp p$, and $r, \tau, G = \mathcal{O}(1)$), this condition reduces to requiring that $\sigma_{\min}(\Factor \Loading_{\gT}^\top)$ scales on the order of $\sqrt{n} + \sqrt{|\gT|}$, which aligns with the standard assumption in spectral analysis. 
The proof of Theorem~\ref{thm:shared_feature_set} is provided in Supplementary Material C.2.

\begin{remarknum}[Analysis of Global PCA Error] \label{rmk:global_alignmnt_error}
 The error bound in~\eqref{eq:bound_shared_feature_set} consists of two terms.
    The first term aggregates the groupwise subspace recovery errors from Stage~1, characterized by $\mathcal{E}_g  / \sigma_{\min}(\Loading_{\gV_{(g)}})$ in Theorem~\ref{thm:shared_feature_set_step_1}. 
    The second term corresponds to the error incurred during the global PCA in Stage~2. 
    Regarding the global PCA error, one may be concerned that performing PCA on the shared anchor set involves a smaller number of features.
    This is reflected in the denominator $\sigma_{\min}(\Loading_{\gT})$, which is typically smaller than
    $\sigma_{\min}(\Loading_{\gV_{(g)}})$ due to the reduced dimension of the anchor set.
    However, unlike the groupwise PCA applied directly to the raw data submatrices in Stage~1, the global PCA is applied to the \emph{projected} submatrix $\mP \rmX_{\cdot, \gT}$. 
    The projection $\mP$ restricts the noise to a low-dimensional subspace, which leads to a smaller numerator $\Xi(\gT)$. In particular, the first term in $\Xi(\gT)$ scales as $r^2 \tau \sqrt{Gr}$, which is substantially smaller than the dominant $\mathcal{O}(\sqrt{n_g})$ term in the groupwise PCA error $\mathcal{E}_g$. 
\end{remarknum}

\begin{remarknum}[Error Bound Comparison]
Following the motivating example in Section~\ref{subsec:challenge}, we consider a regime where the shared feature block and the groupwise subject embeddings may both be weak, while the groupwise combined feature blocks remain strong. 
Specifically, assume balanced block and group sizes with $n_g \asymp n$ and $p_{g} \asymp p$, and suppose the rank, noise level, and number of groups satisfy $\tau, r, G = \mathcal{O}(1)$. 
We further assume the following signal strength:
$\sigma_{\min}(\Factor_{\gU_g}) \asymp n^{\alpha_g/2}$,
$\sigma_{\min}(\Loading_{\gV_{(g)}}) \asymp p^{1/2}$, and
$\sigma_{\min}(\Loading_{\gT}) \asymp p^{\beta/2}$ with some $\alpha_g, \beta \in (0,1]$. 
Under this regime, the global error bound in~\eqref{eq:bound_shared_feature_set} simplifies to 
\begin{equation*}
        \frac{1}{\sqrt{n}} \|\Delta {\widehat{\Factor}}\|_F \lesssim \frac{1}{\sqrt{p}} + \frac{1}{\sqrt{n^{1+\min_g \alpha_g}}}\left(1 + \frac{n}{p}\right) + \frac{\log n}{\sqrt{np^\beta}} + \frac{p}{n p^\beta},
    \end{equation*}
which implies the informal rate in~\eqref{eq:projected_pca_bound_toy_example}.
When $n \asymp p$, it further reduces to $\mathcal{O}(p^{-1/2}+p^{-\beta})$. Compared with using only the shared feature block, which yields the error rate $\mathcal{O}(p^{-\beta/2})$ in~\eqref{eq:bai_ng_bound}, we improve the estimation error bound by leveraging strong signals in $\gV_{(g)}$, while the impact of weak subject embeddings via $\alpha_g$ appears only in smaller-order term $n^{-(1+\min_g \alpha_g)/2}(1+n/p)$, and is therefore negligible under $n \asymp p$.
\end{remarknum}

\subsection{Estimation Error Bound for Algorithm~\ref{alg:projected_pca_chain_link}}

Algorithm~\ref{alg:projected_pca_chain_link} integrates the subspace estimates $\{\tilde{\Factor}_{\gS_k}\}_{k=1}^K$ for all super-groups via a sequential alignment procedure. Therefore, the overall estimation error for the global subject embedding~$\Factor$ can be decomposed into two sources: (1) the estimation errors incurred within each super-group, and (2) the alignment errors propagated along the sequential chain~linking.

For a given super-group $\gS_k$ ($1 \le k \le K$), let $\gT_k \coloneqq \bigcap_{g: \gU_g \subseteq \gS_k} \gV_{(g)}$ denote the set of feature blocks that are observed by all subjects in $\gS_k$, which serves as the \textit{inner anchor}.
Leveraging the results in Theorem~\ref{thm:shared_feature_set}, we define the \textit{super-group-wise estimation error bound}  for the estimator $\tilde{\Factor}_{\gS_k}$ as:
\begin{equation} \label{eq:def_Gamma}
    \Gamma_k \coloneqq \sqrt{ \sum_{g: \gU_g \subseteq \gS_k} \left( \frac{\mathcal{E}_{g}}{\sigma_{\min}(\Loading_{\gV_{(g)}})}\right)^2 } + \frac{\Xi(\gT_k; \gS_k) + \log n}{\sigma_{\min}(\Loading_{\gT_k})},
\end{equation}
where $\mathcal{E}_g$ is defined in~\eqref{eq:local_subspace_recovery_error_bound}, and $\Xi(\gT_k; \gS_k)$ denotes the PCA error associated with the super-group inner anchor block $\rmX_{\gS_k, \gT_k}$. Here, the notation $\Xi(\gT_k; \gS_k)$ is adapted from~\eqref{eq:def_Xi} to the super-group setting by replacing $G$ with the number of groups in $\gS_k$ and $\Factor$ with the super-group-specific subject embeddings $\Factor_{\gS_k}$. By Theorem~\ref{thm:shared_feature_set}, the estimation error for $\tilde{\Factor}_{\gS_k}$ satisfies $\|\Delta\hat{\Factor}_{\gS_k}\|_F \lesssim \sqrt{r}\; \Gamma_k$ with high probability.
The following theorem characterizes how these super-group-wise estimation errors propagate through the chain linking process. Its proof is provided in Supplementary Material C.3.

\begin{theorem} \label{thm:chain_link}
Assume the model in \eqref{eq:model}. For steps $k = 2, \dots, K$ of Algorithm~\ref{alg:projected_pca_chain_link}, let $\gO_k = \left( \bigcup_{j=1}^{k-1} \gS_j \right) \cap \gS_k$ denote the set of overlapping subjects and $\gR_k = \gS_k \setminus \left( \bigcup_{j=1}^{k-1} \gS_j \right)$ denote the set of new subjects. 
Suppose there exists a constant $C_1>0$ such that 
\begin{enumerate}
    \item Condition in~\eqref{eq:assump_local_snr} holds for all groups $g=1\dots, G$;
    \item Condition in~\eqref{eq:assump_global_snr} holds for all super-groups $\gS_k$, $k=1, \dots, K$;
    \item the signal strength on each overlap satisfies $\sigma_{\min}(\Factor_{\gO_k}) \geq C_1 \; \sqrt{r} \; \Gamma_k$, for $k=2, \dots, K$. 
\end{enumerate}
Then there exist constants $c$ and $C_2>0$ such that,  with probability at least $1 - G(G+3)(\min_g n_g)^{-c}$, the output $\widehat{\Factor}$ of Algorithm~\ref{alg:projected_pca_chain_link} satisfies:
\begin{equation}\label{eq:algo2 error bound}
    \frac{1}{\sqrt{n}}\|\Delta {\widehat{\Factor}}\|_F \leq C_2 \; \frac{\sqrt{r}}{\sqrt{n}} \; \sum_{k=1}^K \left( \prod_{k'=k}^K (1 + \omega_{k'}) \right)  \Gamma_k,
\end{equation}
where $\Gamma_k$ is defined in~\eqref{eq:def_Gamma}, and $\omega_k$ denotes the \textit{error amplification factor} at step $k$, with $\omega_1 = 0$ and $\omega_k = \sigma_{\max}(\Factor_{\gR_k})/\sigma_{\min}(\Factor_{\gO_k})$  for $k=2, \dots, K$.
\end{theorem}

\begin{remarknum}
    The bound in~\eqref{eq:algo2 error bound} suggests multiplicative error propagation that arises in sequential alignment procedures. In particular, the super-group-wise estimation error $\Gamma_k$ is amplified by the multiplier $\prod_{k'=k}^K (1+\omega_{k'})$, a phenomenon also observed in sequential alignment procedures for matrix completion~\citep{Zheng2025-fl} and cluster quilting \citep{Zheng2024-ud}. 
    Despite this common amplification effect, our method achieves robustness to weak subject signals by substantially reducing each super-group-wise estimation errors $\Gamma_k$. Specifically, our method leverages Algorithm~\ref{alg:projected_pca_full_observe} to recover the super-group subject embeddings $\Factor_{\gS_k}$. As established in Theorem~\ref{thm:shared_feature_set}, this procedure leads a more accurate estimator $\tilde{\Factor}_{\gS_k}$ via anchor projection. Consequently, although multiplicative error amplification is unavoidable in sequential alignment, its overall impact is mitigated by shrinking the super-group estimation error prior to propagation through the alignment chain.
\end{remarknum}

\subsection{New Representation Reconstruction Error Bound} \label{subsec: spectral slicing}
As noted earlier, the error bound for groupwise subject representations in Theorem~\ref{thm:shared_feature_set_step_1} is robust to the heterogeneous signal strength of subject embeddings. This result follows from a new reconstruction error bound that is robust to heterogeneous signal strengths. We present this bound here as a result of independent interest. 

Consider a general low-rank representation learning problem. Let $\rmX_* = \Factor_* \Loading_*^\top + \rmE_*$, where $\Factor_* \in \R^{n_* \times r}$, $\Loading_* \in \R^{p_* \times r}$, and $\rmE_* \in \R^{n_* \times p_*}$ is an error matrix. 
Let ${\U}_*$ and $\hat{\U}_*$ denote the top-$r$ left singular vectors of the signal matrix $\mA_* = \Factor_*\Loading_*^\top$ and the observed matrix $\rmX_*$, respectively, with associated projection matrices  $\mP_{{\U}_*}$ and $\mP_{\hat{\U}_*}$.
By~\eqref{eq:embedding_subspace_equivalence}, the reconstruction error for $\hat{\Factor}_*=\sqrt{n_*} \hat{\U}_*$ satisfies $\|\Delta \hat{\Factor}_*\|_{F}=\|(\mI - \mP_{\hat{\U}_*})\Factor_*\|_{F} = \|(\mP_{{\U_*}}-\mP_{\hat{\U}_*})\Factor_*\|_{F}$, thereby the reconstruction accuracy is controlled by the subspace estimation error.
A standard approach is to first bound the subspace error via Wedin’s theorem~\citep{Wedin1972-cu} and then convert it to a reconstruction bound:
\begin{equation}\label{eq:naive bound}
    \big\|(\mI-\mP_{\hat{\U}_*})\Factor_*\big\|_{F}
    \lesssim
    \sqrt{r}\,\sigma_{\max}(\Factor_*)\cdot \frac{\|\rmE_*\|}{\sigma_{\min}(\mA_*)}
    \lesssim
    \sqrt{r}\,\kappa(\Factor_*)\cdot \frac{\|\rmE_*\|}{\sigma_{\min}(\Loading_*)},
\end{equation}
where $\kappa(\Factor_*)$ denotes the condition number of $\Factor_*$. The dependence on $\kappa(\Factor_*)$ is undesirable when the subject embeddings are weak, as $\kappa(\Factor_*)$ may be large. 
However, this issue is intrinsic when blockwise missingness meets heterogeneous signal strengths, which motivates a reconstruction bound that remains sharp under weak signal strengths.

\begin{restatable}{proposition}{PropSpectralSlicing} \label{prop:spectral_slicing}
    Suppose the signal matrix $\mA_* = \Factor_*\Loading_*^\top$ satisfies $\sigma_{\min}(\mA_*) \ge 2\max\left\{ 64\, r \, \|\rmE_*\V_*\|,\;  4\, \sqrt{r} \,\|\rmE_*\|\right\},$
    where $\V_* \in \R^{p \times r}$ denote the right $r$ orthonormal singular vectors of $\mA_*$.
    Then there exists a universal constant $C>0$ such that
    \begin{equation*}
        \frac{1}{\sqrt{n}}\left\|(\mI - \mP_{\hat{\U}_*})\Factor_*\right\|_{F}
        \leq
        \frac{Cr^2\sqrt{r}}{\sqrt{n}\,\sigma_{\min}(\Loading_*)}
        \left(
            \|\rmE_*\V_*\|
            +
            \frac{\|\rmE_*\rmE_*^\top\|}{\sigma_{\min}(\mA_*)}
        \right).
    \end{equation*}
\end{restatable}

Compared to~\eqref{eq:naive bound},  Proposition~\ref{prop:spectral_slicing} yields
\begin{equation*}
    \big\|(\mI-\mP_{\hat{\U}_*})\Factor_*\big\|_{F}
    \lesssim
    r^2\sqrt{r}
    \left(1+\frac{\|\rmE_*\|}{\sigma_{\min}(\mA_*)}\right)
    \cdot \frac{\|\rmE_*\|}{\sigma_{\min}(\Loading_*)}.
\end{equation*}
When $r=\mathcal{O}(1)$, the potentially large factor $\kappa(\Factor_*)$ is eliminated and replaced by
$1+|\rmE_*|/\sigma_{\min}(\mA_*)$, which is $\mathcal{O}(1)$ under the signal strength condition
$\sigma_{\min}(\mA_*)\gtrsim \|\rmE_*\|$. In other words, the reconstruction error
remains well controlled even when $\Factor_*$ is ill-conditioned, provided the \emph{product} signal $\mA_*=\Factor_*\Loading_*^\top$ is sufficiently separated from the noise.

The proof of Proposition~\ref{prop:spectral_slicing}, given in Supplementary Material B.1, is based on a \emph{spectral slicing} technique.
Rather than relying on a single global spectral gap, we partition the $r$ singular values of $\mA_*$ into contiguous slices $\gI_1,\dots,\gI_K$ such that within each slice the singular values are of comparable scale, while adjacent slices are
separated by large gaps.
Let $\U_{\gI_k}$ denote the left singular vectors of $\mA_*$ associated with the singular values  in $\gI_k$.
We decompose the reconstruction error as:
\[
    \|(\mI-\mP_{\hat{\U}_*})\Factor_*\|_{F}
    ~\le~
    \sum_{k=1}^K
    \|(\mI-\mP_{\hat{\U}_*})\,\mP_{\U_{\gI_k}}\,\Factor_*\|_{F}.
\]
By applying a sharpened perturbation bound on each slice~$\gI_k$, we leverage a crucial trade-off: 
slices with larger singular values contribute more via $\|\mP_{\U_{\gI_k}}\Factor_*\|_{F}$ but benefit from more accurate subspace estimation,
whereas slices with smaller singular values have less impact on the total error but suffer from less accurate subspace estimation.
This ``cancellation'' across spectral slices yields a bound that remains stable even for poorly conditioned $\Factor_*$.

Proposition~\ref{prop:spectral_slicing} provides the main technical tool for analyzing groupwise subject embeddings in our problem setup. For example, Theorem~\ref{thm:shared_feature_set_step_1} follows by  applying Proposition~\ref{prop:spectral_slicing} with $\rmX_*=\rmX_{\gU_g,\gV_{(g)}}$, $\Factor_*=\Factor_{\gU_g}$, and $ \Loading_*=\Loading_{\gV_{(g)}}$.

\section{Simulation Studies} \label{sec:simulation}

We conduct simulation studies to assess the empirical performance of the proposed methods under blockwise missingness.
We consider two  block structures: a $2\times3$ setting with a shared feature block, and a $3\times3$ setting with no globally shared block.
These correspond to the blue dashed frame and red dotted frame in Figure~\ref{fig:illustration_blockwise_missing}, respectively.

\noindent \textbf{Simulation Setup.} We generate data following the model~\eqref{eq:model} with rank $r=6$ and noise $\rmE_{ij}\stackrel{\text{i.i.d.}}{\sim}\mathcal{N}(0,1)$.
Initial subject embeddings $\Factor_{\gU_g}^\dagger\in\mathbb{R}^{n\times r}$ and feature embeddings
$\Loading_{\gV_b}^\dagger\in\mathbb{R}^{p\times r}$ are drawn independently with i.i.d.\ $\mathcal{N}(0,1)$ entries.
We impose blockwise signal heterogeneity by diagonal rescaling of $r$ coordinates. Define $s:=n^{(\alpha-1)/2}$ and $t:=p^{(\beta-1)/2}$.
\begin{itemize}
    \item $2\times3$ setting: We consider two subject groups $g\in\{1,2\}$ and three feature sets $b\in\{1,2,3\}$, where set~1 is shared across groups.
We set complementary signal strengths in two subject groups:
$\Factor_{\gU_1}=\Factor_{\gU_1}^\dagger\operatorname{diag}(s,s,s,1,1,1)$ and
$\Factor_{\gU_2}=\Factor_{\gU_2}^\dagger\operatorname{diag}(1,1,1,s,s,s)$.
For the feature signal strengths, we set
$\Loading_{\gV_1}=t\,\Loading_{\gV_1}^\dagger$, $\Loading_{\gV_2}=\Loading_{\gV_2}^\dagger$, and
$\Loading_{\gV_3}=\Loading_{\gV_3}^\dagger$, so~that the shared feature signal is weak while the aggregated feature signal remains strong. 
With high probability, this construction yields the signal strength~in~\eqref{eq:toy_example_def_beta}–\eqref{eq:toy_example_def_alpha}.
\item $3\times3$ setting: 
We consider three subject groups $g\in\{1,2,3\}$ and three feature sets $b\in\{1,2,3\}$, with no feature set observed across all groups.
We impose complementary signal strengths across subject groups and feature sets:
$\Factor_{\gU_1}=\Factor_{\gU_1}^\dagger\operatorname{diag}(s,s,1,1,1,1)$,
$\Factor_{\gU_2}=\Factor_{\gU_2}^\dagger\operatorname{diag}(1,1,s,s,1,1)$,
$\Factor_{\gU_3}=\Factor_{\gU_3}^\dagger\operatorname{diag}(1,1,1,1,s,s)$, 
$\Loading_{\gV_1}=\Loading_{\gV_1}^\dagger\operatorname{diag}(t,t,1,1,1,1)$,
$\Loading_{\gV_2}=\Loading_{\gV_2}^\dagger\operatorname{diag}(1,1,t,t,1,1)$,
and $\Loading_{\gV_3}=\Loading_{\gV_3}^\dagger\operatorname{diag}(1,1,1,1,t,t)$.
This construction yields $\sigma_{\min}(\Factor_{\gU_g})\asymp n^{\alpha/2}$ and $\sigma_{\min}(\Loading_{\gV_b})\asymp p^{\beta/2}$ for each $g,b$, while any pair of subject groups (or feature blocks) jointly exhibits strong aggregated signal.

\end{itemize}

Throughout, we fix $\alpha=\beta=0.5$. 
Across all settings, we evaluate recovery of the subject representations and report the normalized estimation error
$\|\Delta \widehat{\Factor}\|_F / \|\Factor\|_F$.

\subsection{Results for Algorithm~\ref{alg:projected_pca_full_observe} under the $2\times3$ Setting}
We compare APPCA in Algorithm~\ref{alg:projected_pca_full_observe} with an oracle estimator (PCA on the fully observed data matrix) and two feasible baselines: PCA on the shared block and a two-step embedding alignment method.

\noindent \textbf{Balanced Growth with $n \asymp p$.}
Figure~\ref{fig:sim_2_3_sim} reports the estimation error as the sample size and feature dimension increase proportionally ($n = p$). APPCA achieves an error~rate~that closely tracks the oracle estimator, which decays at order $\mathcal{{O}}(p^{-1/2})$, indicating effective aggregation of information across all observed blocks. In contrast, PCA applied to the shared block exhibits a slower convergence rate, as its performance is bottlenecked~by the weak signal in the shared feature set. The two-step embedding alignment method also~underperforms, which can be attributed to inaccurate feature embedding estimation when subject signals are weak in group-specific blocks, which in turn leads to poor alignment~in the second stage.

\begin{figure}[!htbp]
    \centering
    \includegraphics[width=0.65\linewidth]{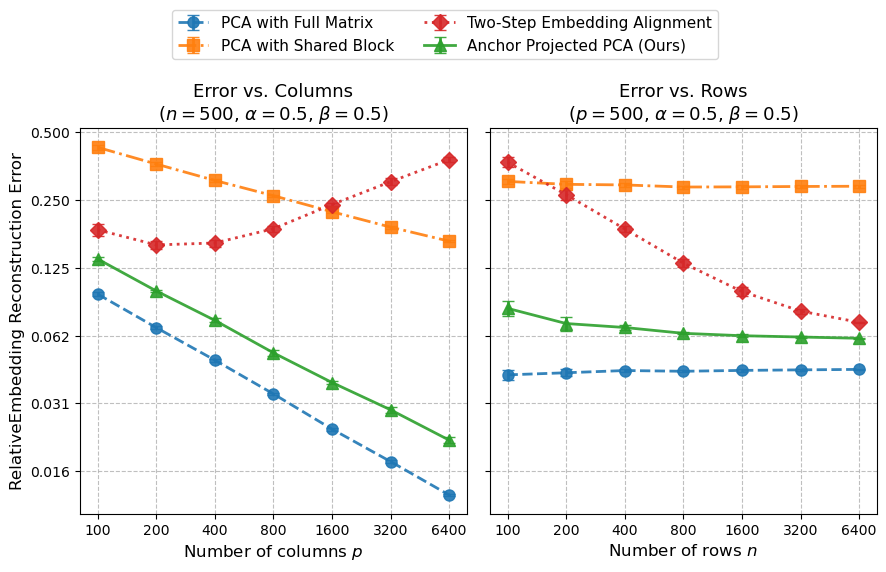}
    \caption{Log-log plots of the estimation error of $\hat{\Factor}$ under the $2\times3$ setting.
Left: \emph{fixed-sample-size regime} with $n=500$ and increasing feature dimension $p$.
Right: \emph{fixed-dimension regime} with $p=500$ and increasing sample size $n$.}
    \label{fig:sim_2_3_fix}
\end{figure}

\noindent \textbf{Varying Feature Size $p$ with Fixed $n=500$.}
As shown in the left panel of Figure~\ref{fig:sim_2_3_fix}, APPCA remains substantially more robust to increasing $p$ than the baselines. It maintains faster error decay than PCA on the shared block. 
The two-step embedding alignment method initially improves but deteriorates rapidly once $p \gg n$, which suggests increasing instability in feature embedding estimation under high dimensionality.

\noindent \textbf{Varying Sample Size $n$ with Fixed $p = 500$.}
When $p$ is fixed, the estimation error is lower bounded by an irreducible error floor regardless of sample size. As shown in the right panel of Figure~\ref{fig:sim_2_3_fix}, all methods exhibit a flattening trend as $n$ grows. Nonetheless, APPCA consistently outperforms the two baselines.
The two-step embedding alignment method improves with increasing $n$ but remains worse than APPCA, particularly for small $n$, while PCA on the shared block shows little sensitivity to $n$, consistent with being bottlenecked by the weak shared feature set.

\subsection{Results for Algorithm~\ref{alg:projected_pca_chain_link} under the $3 \times 3$ Setting}
\begin{figure}[!htbp]
    \centering
    \includegraphics[width=0.95\linewidth]{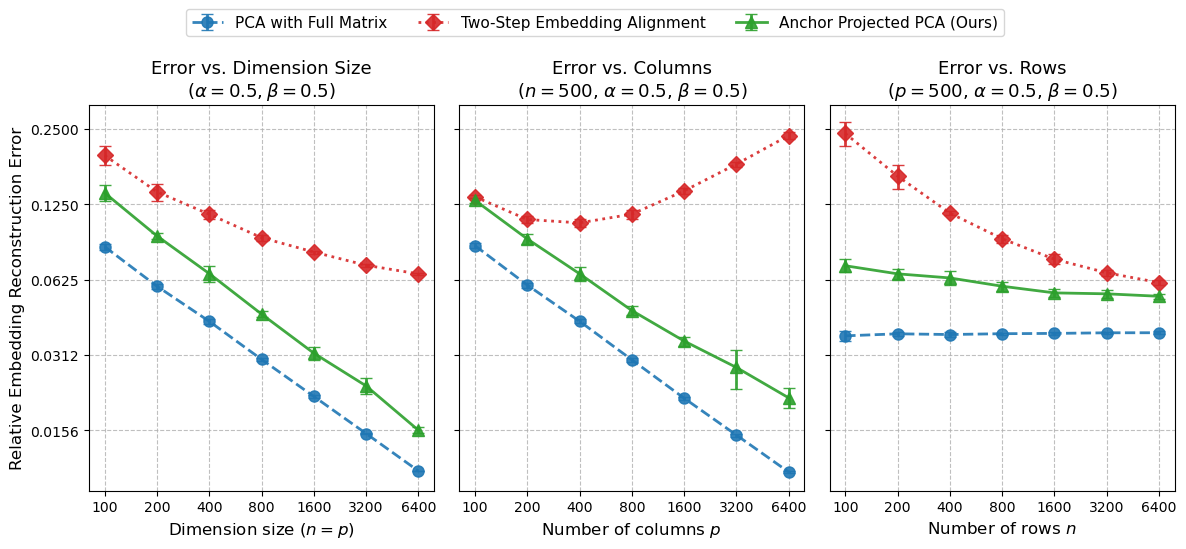}
    \caption{Log-log plots of the estimation error of $\widehat{\Factor}$ under the $3\times3$ setting. Left: \emph{balanced-growth regime} with $n \asymp p$.
Middle: \emph{fixed-sample-size regime} with $n=500$ and increasing $p$.
Right: \emph{fixed-dimension regime} with $p=500$ and increasing $n$.}
    \label{fig:sim_3_3}
\end{figure}

We next evaluate Double APPCA in Algorithm~\ref{alg:projected_pca_chain_link} and compare it with the two-step embedding alignment baseline. The shared-block baseline is omitted as no globally shared feature block exists.
Figure~\ref{fig:sim_3_3} summarizes the results under the balanced-growth, fixed-sample-size, and fixed-feature-dimension regimes. Across all regimes, Double APPCA consistently outperforms the two-step embedding alignment method. Although the chain-linking procedure introduces additional error accumulation relative to the $2\times3$ setting, the overall performance trends closely mirror those observed in the $2\times3$ case (Algorithm~\ref{alg:projected_pca_full_observe}). 

\section{Real-World Data Example} \label{sec:real_data}

In this section, we apply the proposed method to a single-cell multimodal dataset. 
Understanding cellular identity and function from multimodal single-cell measurements is an important problem in modern genomics~\citep{Hao2021-bs, Ghazanfar2024-jy}. Different molecular modalities capture varying aspects of cellular state, and integrating these sources of information is essential for accurate cell characterization and downstream biological discovery. In practice, however, multimodal data integration is challenging. First, different experimental platforms often measure different subsets of modalities, leading to structured blockwise missingness. Second, datasets generated using different protocols often exhibit substantial biological heterogeneity, such as shifts in cell-type composition or activation states, further complicating integration. These challenges motivate the need for representation learning methods that can recover shared latent cellular structure from partially observed and heterogeneous multimodal data~\citep{Luecken2021-ea}.
Our goal here is to evaluate and compare our proposed APPCA with existing methods for learning cell representations under blockwise missingness and heterogeneity.

We use a single-cell multimodal dataset generated via TEA-seq~\citep{Swanson2021-ky}\footnote{The complete dataset is publicly available under the Gene Expression Omnibus accession number GSE158013}, which simultaneously profiles RNA gene expression, chromatin accessibility (ATAC), and cell surface protein abundance (ADT) for peripheral blood mononuclear cells (PBMCs).
This fully observed dataset enables us to impose controlled blockwise missingness patterns that mirror common experimental protocols while still allowing quantitative evaluation of the learned representations.
To this end, we construct two subject groups with disjoint missing modalities. 
We first randomly subsample $800$ Naive T cells and $300$ Memory T cells from the original data and assign them unevenly to the two groups, inducing heterogeneity in cell-type composition. Group~1 consists of 95\% Naive T cells and 5\% Memory T cells, while Group~2 contains the remaining cells, resulting in Naive-dominant and Memory-dominant populations, respectively (see Table~\ref{tab:missingness}).
In Group~1, we retain RNA and ADT measurements while masking ATAC features, mimicking surface-protein-based assays such as CITE-seq. In Group~2, we retain RNA and ATAC measurements while masking ADT features, mimicking joint RNA–chromatin assays such as 10x Multiome. In both groups, RNA serves as the shared modality linking the datasets, while the other modalities are mutually exclusive, reflecting a common mosaic integration setting~\citep{Ghazanfar2024-jy, Luecken2021-ea}.
For feature selection, we keep the top $600$ RNA and $500$ ATAC features identified as highly variable using the standard Scanpy pipeline~\citep{Wolf2018-nr}. 
Additional details on data preprocessing are provided in Supplementary Material E. 

\begin{table}[!htbp]
\centering
\caption{Summary of the experimental design. The RNA view serves as the shared bridge connecting the Naive-dominant and Memory-dominant populations.}
\label{tab:missingness}
\begin{tabular}{lcccc}
\toprule
\textbf{Subject Group} & \textbf{State Composition} & \textbf{RNA} & \textbf{ADT} & \textbf{ATAC} \\
 & & ($p_1=600$) & ($p_2=47$) & ($p_3=500$) \\
\midrule
Group 1 $(n_1 = 775)$ & Naive-Dominant & $\checkmark$ & $\checkmark$ & \textit{Missing} \\
Group 2 $(n_2 = 325)$ & Memory-Dominant & $\checkmark$ & \textit{Missing} & $\checkmark$ \\
\bottomrule
\end{tabular}
\end{table}

To evaluate our method, we compare the learned representation $\widehat{\Factor}$ with an oracle embedding, $\Factor_{\mathrm{oracle}}$, derived from the complete dataset. Specifically, we define the oracle embedding as $\Factor_{\mathrm{oracle}} = \sqrt{n} \cdot \U_{\mathrm{oracle}}$, where $\U_{\mathrm{oracle}}$ consists of the top-$r$ left singular vectors of the full data matrix and $n = n_1 + n_2$ denotes the total number of cells. We measure the discrepancy between the learned and oracle embeddings as $\left\| \Delta \widehat{\Factor} \right\|_F / \left\|\Factor_{\mathrm{oracle}} \right\|_F$,
which evaluates alignment up to a linear transformation.
This metric quantifies the ability to recover the latent representation that would have been obtained from a fully observed~dataset.

We compare APPCA with the two baseline strategies introduced in Section~\ref{sec:problem_setup}: PCA applied to the shared block and the two-step embedding alignment approach. For all methods, the rank is selected as $r=9$ using the information criterion proposed in~\citet{Bai2002-zy}. We use the cross-fit variant of APPCA and report the average reconstruction error along with its standard deviation across $200$ independent random two-fold splits. For the two-step alignment, which is order-dependent, we present results for both alignment directions (mapping Group 2 to Group 1, and vice versa). We summarize the results in Table~\ref{tab:results}. 

\begin{table}[!htbp]
\centering
\caption{Performance comparison on the TEA-seq dataset. For APPCA, the result is reported as Mean (Standard Deviation) over 200 random splits. Baselines are deterministic.}
\label{tab:results}
\begin{tabular}{lc}
\toprule
\textbf{Method} & \textbf{Reconstruction Error} ($\downarrow$) \\
\midrule
PCA (Shared Block) & 0.532 \\
Two-Step Alignment (Group 1 $\rightarrow$ Group 2) & 0.521 \\
Two-Step Alignment (Group 2 $\rightarrow$ Group 1) & 0.605 \\
\textbf{APPCA} & \textbf{0.503} (0.018) \\
\bottomrule
\end{tabular}
\end{table}

APPCA achieves the lowest reconstruction error among all compared methods. 
PCA applied only to the shared block yields a higher error, indicating that APPCA effectively borrows information from the partially observed ATAC and ADT modalities to refine the underlying cellular states, rather than relying solely on the shared RNA view.
Importantly, the results also reveal a limitation of two-step embedding alignment methods when applied to asymmetric data matrices. Their performance is highly sensitive to the direction of embedding alignment: mapping Group~1 to Group~2 achieves an error of 0.521, whereas reversing the direction substantially degrades performance to 0.605, even worse than Shared Block PCA. This difference suggests that feature-embedding-based alignment strategies are sensitive to the quality of the reference group, with the choice of reference substantially affecting the final globally aligned embedding. In contrast, APPCA performs joint subspace recovery on projected data matrices and avoids arbitrary reference choices. 

\section{Discussion}

In this work, we proposed APPCA, a representation learning framework designed for multi-source integration under two critical challenges: structured blockwise missingness and blockwise signal heterogeneity. 
APPCA adopts a two-stage strategy that first performs robust groupwise subspace recovery using all available feature blocks within each group, and then derives globally aligned representations based on the projected anchor block. By projecting the shared anchor block onto reliably estimated low-dimensional subspaces, APPCA effectively aggregates information across heterogeneous sources while suppressing high-dimensional noise, even when the anchor signal itself is weak.

On the theoretical side, we developed a fine-grained perturbation analysis based on a new spectral slicing technique. 
This analysis yields reconstruction error bounds that remain well controlled under heterogeneous and weak signal regimes and, crucially, avoid the standard dependence on the condition number of subject embeddings. These results provide theoretical insight into how robustness is achieved at each stage of APPCA and explain its strong empirical performance in challenging multi-source settings.

In the chain-linked extension of APPCA, multiple merging sequences may satisfy the connectivity conditions in Section~\ref{subsec:chain_link}. Our theory highlights a trade-off in this choice. Coarser chains shorten the alignment path and reduce error accumulation, while finer chains retain stronger shared signal but may incur larger cumulative error due to longer chains. Characterizing this trade-off more precisely and developing data-driven strategies for selecting effective chain-linking sequences remain important directions for future work.

\spacingset{1.8} 
\bibliography{reference}

\begin{thebibliography}{32}
\providecommand{\natexlab}[1]{#1}
\providecommand{\url}[1]{\texttt{#1}}
\providecommand{\urlprefix}{URL }
\expandafter\ifx\csname urlstyle\endcsname\relax
  \providecommand{\doi}[1]{doi:\discretionary{}{}{}#1}\else
  \providecommand{\doi}{doi:\discretionary{}{}{}\begingroup \urlstyle{rm}\Url}\fi
\providecommand{\eprint}[2][]{\url{#2}}

\bibitem[{Bai and Ng(2002)}]{Bai2002-zy}
Bai, J. and Ng, S.
\newblock Determining the number of factors in approximate factor models.
\newblock \emph{Econometrica: Journal of the Econometric Society}, 70(1):191--221, 2002.

\bibitem[{Bai and Ng(2023)}]{Bai2023-hj}
Bai, J. and Ng, S.
\newblock Approximate factor models with weaker loadings.
\newblock \emph{Journal of Econometrics}, 235(2):1893--1916, 2023.

\bibitem[{Bishop and Yu(2014)}]{Bishop2014-zj}
Bishop, W.~E. and Yu, B.~M.
\newblock Deterministic symmetric positive semidefinite matrix completion.
\newblock \emph{Neural Information Processing Systems}, 27:2762--2770, 2014.

\bibitem[{Cai et~al.(2016)Cai, Cai, and Zhang}]{Cai2016-vw}
Cai, T., Cai, T.~T., and Zhang, A.
\newblock Structured matrix completion with applications to genomic data integration.
\newblock \emph{Journal of the American Statistical Association}, 111(514):621--633, 2016.

\bibitem[{Chang et~al.(2024)Chang, Russo, and Paul}]{Chang2024-bq}
Chang, J.~H., Russo, M., and Paul, S.
\newblock Heterogeneous transfer learning for high dimensional regression with feature mismatch.
\newblock \emph{arXiv [stat.ML]}, 2024.

\bibitem[{Choi and Yuan(2025)}]{Choi2025-lp}
Choi, J. and Yuan, M.
\newblock High dimensional factor analysis with weak factors.
\newblock \emph{Journal of Econometrics}, 252(106086):106086, 2025.

\bibitem[{Fan and Liao(2022)}]{Fan2022-kf}
Fan, J. and Liao, Y.
\newblock Learning latent factors from diversified projections and its applications to over-estimated and weak factors.
\newblock \emph{Journal of the American Statistical Association}, 117(538):909--924, 2022.

\bibitem[{Fan et~al.(2016)Fan, Liao, and Wang}]{Fan2016-af}
Fan, J., Liao, Y., and Wang, W.
\newblock Projected principal component analysis in factor models.
\newblock \emph{The Annals of Statistics}, 44(1):219--254, 2016.

\bibitem[{Freyaldenhoven(2022)}]{Freyaldenhoven2022-yj}
Freyaldenhoven, S.
\newblock Factor models with local factors -- determining the number of relevant factors.
\newblock \emph{Journal of Econometrics}, 229(1):80--102, 2022.

\bibitem[{Ghazanfar et~al.(2024)Ghazanfar, Guibentif, and Marioni}]{Ghazanfar2024-jy}
Ghazanfar, S., Guibentif, C., and Marioni, J.~C.
\newblock Stabilized mosaic single-cell data integration using unshared features.
\newblock \emph{Nature Biotechnology}, 42(2):284--292, 2024.

\bibitem[{Hao et~al.(2021)Hao, Hao, Andersen-Nissen et~al.}]{Hao2021-bs}
Hao, Y., Hao, S., Andersen-Nissen, E., et~al.
\newblock Integrated analysis of multimodal single-cell data.
\newblock \emph{Cell}, 184(13):3573--3587.e29, 2021.

\bibitem[{Hripcsak and Albers(2013)}]{Hripcsak2013-hq}
Hripcsak, G. and Albers, D.~J.
\newblock Next-generation phenotyping of electronic health records.
\newblock \emph{Journal of the American Medical Informatics Association}, 20(1):117--121, 2013.

\bibitem[{Jin and Rothenh{\"{a}}usler(2023)}]{Jin2023-df}
Jin, Y. and Rothenh{\"{a}}usler, D.
\newblock Modular regression: Improving linear models by incorporating auxiliary data.
\newblock \emph{Journal of Machine Learning Research}, 24(351):1--52, 2023.

\bibitem[{Kundu et~al.(2019)Kundu, Tang, and Chatterjee}]{Kundu2019-ku}
Kundu, P., Tang, R., and Chatterjee, N.
\newblock Generalized meta-analysis for multiple regression models across studies with disparate covariate information.
\newblock \emph{Biometrika}, 106(3):567--585, 2019.

\bibitem[{Luecken et~al.(2021)Luecken, Burkhardt, Cannoodt et~al.}]{Luecken2021-ea}
Luecken, M., Burkhardt, D., Cannoodt, R., et~al.
\newblock A sandbox for prediction and integration of {DNA}, {RNA}, and proteins in single cells.
\newblock In J.~Vanschoren and S.~Yeung, editors, \emph{Proceedings of the Neural Information Processing Systems Track on Datasets and Benchmarks}, volume~1. 2021.

\bibitem[{Onatski(2012)}]{Onatski2012-xz}
Onatski, A.
\newblock Asymptotics of the principal components estimator of large factor models with weakly influential factors.
\newblock \emph{Journal of Econometrics}, 168(2):244--258, 2012.

\bibitem[{Shi et~al.(2021)Shi, Li, and Cai}]{Shi2021-bu}
Shi, X., Li, X., and Cai, T.
\newblock Spherical regression under mismatch corruption with application to automated knowledge translation.
\newblock \emph{Journal of the American Statistical Association}, 116(536):1953--1964, 2021.

\bibitem[{Song et~al.(2024)Song, Lin, and Zhou}]{Song2024-er}
Song, S., Lin, Y., and Zhou, Y.
\newblock Semi-supervised inference for block-wise missing data without imputation.
\newblock \emph{Journal of Machine Learning Research}, 25(99):99:1--99:36, 2024.

\bibitem[{Stuart et~al.(2019)Stuart, Butler, Hoffman et~al.}]{Stuart2019-st}
Stuart, T., Butler, A., Hoffman, P., et~al.
\newblock Comprehensive integration of single-cell data.
\newblock \emph{Cell}, 177(7):1888--1902.e21, 2019.

\bibitem[{Sui et~al.(2025)Sui, Xu, Bai et~al.}]{Sui2025-ry}
Sui, Y., Xu, Q., Bai, Y., et~al.
\newblock Multi-task learning for heterogeneous multi-source block-wise missing data.
\newblock \emph{arXiv [cs.LG]}, 2025.

\bibitem[{Swanson et~al.(2021)Swanson, Lord, Reading et~al.}]{Swanson2021-ky}
Swanson, E., Lord, C., Reading, J., et~al.
\newblock Simultaneous trimodal single-cell measurement of transcripts, epitopes, and chromatin accessibility using {TEA}-seq.
\newblock \emph{eLife}, 10:e63632, 2021.

\bibitem[{Uematsu and Yamagata(2023)}]{Uematsu2023-ng}
Uematsu, Y. and Yamagata, T.
\newblock Estimation of sparsity-induced weak factor models.
\newblock \emph{Journal of Business \& Economic Statistics}, 41(1):213--227, 2023.

\bibitem[{Wedin(1972)}]{Wedin1972-cu}
Wedin, P.-A.
\newblock Perturbation bounds in connection with singular value decomposition.
\newblock \emph{BIT numerical mathematics}, 12(1):99--111, 1972.

\bibitem[{Wolf et~al.(2018)Wolf, Angerer, and Theis}]{Wolf2018-nr}
Wolf, F.~A., Angerer, P., and Theis, F.~J.
\newblock {SCANPY}: large-scale single-cell gene expression data analysis.
\newblock \emph{Genome Biology}, 19(1):15, 2018.

\bibitem[{Xu et~al.(2025)Xu, Testa, Lei et~al.}]{Xu2025-jr}
Xu, Q., Testa, L., Lei, J., et~al.
\newblock Blockwise missingness meets {AI}: A tractable solution for semiparametric inference.
\newblock \emph{arXiv [stat.ME]}, 2025.

\bibitem[{Xue et~al.(2021)Xue, Ma, and Li}]{Xue2021-ae}
Xue, F., Ma, R., and Li, H.
\newblock Statistical inference for high-dimensional linear regression with blockwise missing data.
\newblock \emph{arXiv [stat.ME]}, 2021.

\bibitem[{Xue and Qu(2021)}]{Xue2021-sq}
Xue, F. and Qu, A.
\newblock Integrating multisource block-wise missing data in model selection.
\newblock \emph{Journal of the American Statistical Association}, 116(536):1914--1927, 2021.

\bibitem[{Yu et~al.(2020)Yu, Li, Shen et~al.}]{Yu2020-ld}
Yu, G., Li, Q., Shen, D., et~al.
\newblock Optimal sparse linear prediction for block-missing multi-modality data without imputation.
\newblock \emph{Journal of the American Statistical Association}, 115(531):1406--1419, 2020.

\bibitem[{Zhang et~al.(2020)Zhang, Tang, and Qu}]{Zhang2020-zc}
Zhang, Y., Tang, N., and Qu, A.
\newblock Imputed factor regression for high-dimensional block-wise missing data.
\newblock \emph{Statistica Sinica}, 2020.

\bibitem[{Zheng et~al.(2024)Zheng, Chang, and Allen}]{Zheng2024-ud}
Zheng, L., Chang, A., and Allen, G.~I.
\newblock Cluster quilting: Spectral clustering for patchwork learning.
\newblock \emph{arXiv [stat.ME]}, 2024.

\bibitem[{Zheng and Tang(2025)}]{Zheng2025-fl}
Zheng, R. and Tang, M.
\newblock Chain-linked multiple matrix integration via embedding alignment.
\newblock \emph{Journal of the American Statistical Association}, pages 1--24, 2025.

\bibitem[{Zhou et~al.(2023)Zhou, Cai, and Lu}]{Zhou2023-wb}
Zhou, D., Cai, T., and Lu, J.
\newblock Multi-source learning via completion of block-wise overlapping noisy matrices.
\newblock \emph{Journal of Machine Learning Research}, 24(221):1--43, 2023.

\end{thebibliography}

\end{document}